\documentclass[a4paper, 12pt, parskip, toc=bibliography,captions = tableheading]{scrartcl}
\usepackage[T1]{fontenc}
\usepackage[utf8]{inputenc}
\usepackage[english]{babel}
\usepackage{xcolor}
\usepackage[en-US]{datetime2}
\usepackage{fancyhdr}
\usepackage[a4paper]{geometry}
\usepackage{tikz}
\usetikzlibrary{positioning,arrows.meta,calc,fit,backgrounds,shapes.geometric}
\usepackage{csquotes}
\usepackage[backend=biber, style=authoryear-icomp, ibidtracker=false, idemtracker=false, uniquelist=false, uniquename=false, maxcitenames=2, maxbibnames = 25, natbib]{biblatex}
\DefineBibliographyStrings{english}{andothers={et\ al\adddot}}
\usepackage{setspace}
\usepackage[all]{nowidow}

\usepackage{enumitem}
\usepackage[most]{tcolorbox}
\tcbset{sharp corners}  % optional: kein abgerundeter Rahmen

\newtcolorbox{codebookbox}{
  colback=gray!2,
  colframe=gray!60,
  boxrule=0.3pt,
  arc=1pt,
  enhanced,
  breakable,
  before skip=10pt,
  after skip=10pt,
  left=8pt,
  right=8pt,
  top=6pt,
  bottom=6pt
}

\usepackage{mathtools, amsmath, amssymb, amsfonts, nicefrac, booktabs}
\usepackage{hyperref}

\title{Identifying economic narratives in large text corpora\\
An integrated approach using Large Language Models}
\author{Tobias Schmidt, Kai-Robin Lange, Matthias Reccius, Henrik Müller, Michael Roos, Carsten Jentsch}
\date{May 2025}

\addto\extrasenglish{%
}

\begin{document}

{\bfseries Identifying economic narratives in large text corpora -- \\
An integrated approach using Large Language Models}\\[0.3em]
\normalsize

Tobias Schmidt$^1$ (TU Dortmund University), Kai-Robin Lange$^1$ (TU Dortmund University), Matthias Reccius$^1$ (Ruhr-University Bochum), Henrik Müller (TU Dortmund University), Michael Roos (Ruhr-University Bochum) and Carsten Jentsch (TU Dortmund University)
\vspace{-1em}
\section*{Abstract}
\vspace{-1em}
As interest in economic narratives has grown in recent years, so has the number of pipelines dedicated to extracting such narratives from texts. Pipelines often employ a mix of state-of-the-art natural language processing techniques, such as BERT, to tackle this task. While effective on foundational linguistic operations essential for narrative extraction, such models lack the deeper semantic understanding required to distinguish extracting economic narratives from merely conducting classic tasks like Semantic Role Labeling. Instead of relying on complex model pipelines, we evaluate the benefits of Large Language Models (LLMs) by analyzing a corpus of Wall Street Journal and New York Times newspaper articles about inflation. We apply a rigorous narrative definition and compare \texttt{GPT-4o} outputs to gold-standard narratives produced by expert annotators. Our results suggests that \texttt{GPT-4o} is capable of extracting valid economic narratives in a structured format, but still falls short of expert-level performance when handling complex documents and narratives. Given the novelty of LLMs in economic research, we also provide guidance for future work in economics and the social sciences that employs LLMs to pursue similar objectives.

JEL-Code: C18, C55, C87, E70

Key words: economic narratives, natural language processing, large language models

$^1$: equal contribution\\
Acknowledgments: This study is part of a project of the Dortmund Center for data-based Media Analysis (DoCMA) at TU Dortmund University and the Narrative Economic Alliance Ruhr (NEAR) project, supported by the Mercator Research Center Ruhr (MERCUR) with project number Ko-2022-0015. It was also partially funded by the Reality Check incubator project at the Research Center for Trustworthy Data Science and Security.

%-Abstract
\newpage
\normalsize

\section{Introduction}
\label{chap1}      

Macroeconomic policy decisions are not made in a vacuum. More and more, central banks and other institutions recognize the role that public discourse and widely shared narratives play in shaping expectations and thus economic behavior. The central bank communications literature in particular has described the active role of policy makers in explaining their policies to great detail \parencite{Gurkaynak2005, Hansen2017, Hansen2019, Gorodnichenko2023}. But neither fiscal nor monetary policy discourses are top-down processes. People's economic views and expectations are increasingly shaped by the media \parencite{DeFiore2025}. Particularly in times of uncertainty, stories about rising inflation, looming recessions, or job market disruptions spread rapidly and uncontrollably. 

As economic dynamics are driven by the beliefs and expectations of households and firms, monitoring the narratives that dominate public discourse is crucial. Quantifying economic narratives in mass media can be challenging, however, as the amount of information transmitted is vast and it's rhetorical packaging diverse. Many researchers perform qualitative analyses, which require expert evaluations of texts. While such procedures produce reliable results, they are not scalable to large text corpora. Increasing numbers of articles, reports, opinion pieces and even comments on social media are released every day, so purely relying on qualitative research is not feasible when analyzing the narratives that circulate in an economy. Therefore, economists need efficient, quantitative methods of extracting and presenting narratives from texts.

State-of-the-art language models such as BERT \parencite{Bert} and its variations have shown considerable success in tasks like sentiment analysis, named entity recognition, and semantic role labeling, which are seen by many as foundational to the identification of economic narratives. However, while these models excel at narrow linguistic tasks, they often fall short when tasked with the deeper, more nuanced understanding required to distinguish between complex economic narratives and simpler textual structures. Pipeline approaches that mix different methods also introduce compounding error risks due to their multiple processing stages. Hence, traditional NLP models address the scalability problem but lack the integrated language comprehension needed to fully capture the contextual meaning and economic relevance of narratives. 

This gap presents an opportunity for exploring alternative approaches that can offer greater contextual understanding. The advent of Large Language Models (LLMs), with both commercial models such as GPT \parencite{brown2020language}, Claude Sonnet and Google Gemini and open source models like Llama \parencite{Llama3}, has opened up entirely new possibilities for narrative extraction. LLMs have demonstrated remarkable proficiency in a wide array of tasks, excelling especially in those that require a deeper comprehension of language and context \parencite{GPT4oBenchmark}. These models are trained on vast amounts of data and can process complex narratives with a level of understanding closer to that of human readers. But how should economists leverage LLMs to accurately measure complex phenomena such as narratives? And can LLMs match expert annotators in identifying and extracting economic narratives from newspaper articles? 

In this paper, we use a set of complex instructions and expert-labeled examples to evaluate the proficiency of LLMs at extracting economic narratives. We utilize \texttt{GPT-4o} \parencite{openai2024gpt4ocard}, the leading commercial LLM at the time of analysis, for the extraction of inflation narratives from a corpus of Wall Street Journal and New York Times newspaper articles. We further demonstrate how complex concepts like economic narratives can be operationalized for human coding, and then adapted for coding by LLMs. While such models enable new avenues for quantitative economic research, the litmus test for their usefulness lies in the rigorous validation of their outputs. Hence, we compare the results of the LLMs’ analysis against a gold-standard set of narratives, representing the consensus of three trained annotators with domain expertise. A detailed and rigorous annotator codebook was developed through multiple iterations and refinements. We find that expert annotators reach notably different results extracting narratives, pointing to the inherent subjectivity involved in narrative sense-making. To refine and adjust our evaluation of GPTs' performance to the subjectivity of the task, we not only consider the gold-standard data set as a set of optimal answers, but also the deviations made by the expert annotators, which we call \enquote{expected} deviations. We also compare the results attainable through a set of different LLM prompting strategies recently found to be the most effective by research in Artificial Intelligence (AI).
% We further show in detail how we operationalize the very complex notion of an economic narrative by shedding light on our work on a precise, yet complex code book for our annotators that is the result of many iterations and improvements. We demonstrate that the definition of a economic narrative is so complex that, even with a precise code book, multiple expert annotators may reach different results, depending on how the text and the authors intentions are interpreted by each individual. To adjust our interpretation of GPT's performance to this subjective task, we not only consider the gold-standard data set as a set of optimal answers, but also the deviations made by the expert annotators as \enquote{expected} deviations.

The rest of the paper is structured as follows. In \autoref{chap2}, we shed light on previous research that has aimed to define economic narratives or automatically extract them from texts. We also provide a short overview of inflation-related literature, as inflation is the center of our evaluation strategy. In \autoref{chap3}, we introduce the data as well as the LLM we use in our experiments. \autoref{chap4} covers our narrative extraction codebook as well as a description of the process of creating a gold-standard data set. In \autoref{chap5} we explain our prompting strategy and compare it to other approaches used in the field of AI to maximize LLM performance. The results of the experiments are then displayed and interpreted in \autoref{chap6} where we also provide an outlook for future research. Finally, we conclude in \autoref{chap7}.

\section{Related Work}
\label{chap2}

\subsection{Defining economic narratives}
\label{DefiningNarratives}
The field of narrative economics has gained remarkable prominence in recent years \parencite{roos2024}. Only eight years after the well-known paper by \textcite{Shiller17} on this topic, it has become widely accepted in the economic literature that individuals do not simply react to economic data, but interpret the world through personal or publicly shared interpretations of what is going on. Such stories, or narratives, are expected to have the potential to shape personal expectations and even guide collective behavior, when widely shared \parencite{benabou2018narratives, Flynn2022, larsen2019business}. This relationship appears particularly relevant in times of uncertainty, when traditional expectation formation models struggle to capture real-world behavior. As \textcite{king2020radical} argue, under conditions of radical uncertainty, economic agents are unable to maximize utility in the conventional sense. Instead, they turn to shared sense-making structures to reduce complexity and navigate decision-making.

Despite the growing recognition that narratives constitute a promising field of research, the definition of what precisely qualifies as an \enquote{economic narrative} remains fragmented, with researchers highlighting different aspects depending on their analytical objectives. Early contributions, most notably \textcite{Shiller17}, describe economic narratives as broad stories that convey interpretations of economic events, morals, or simplified theories. While Shiller's intuitive approach to narratives captures their communicative power, high-level concepts like \enquote{moral} or \enquote{interpretation} resist the consistent formal structure required for empirical analysis. More recent studies, therefore, aim to articulate definitions that are both theoretically rigorous and operationalizable.

A central contribution to the formalization of economic narratives comes from \textcite{eliaz2020model, eliaz2024news}, who conceptualize narratives as simplified causal models represented by directed acyclic graphs (DAGs). Drawing on Bayesian network theory, their 2020 model shows how individuals adopt competing narrative-policy pairs that interpret long-run correlations to maximize anticipatory utility. Narratives, in their framework, are not neutral representations of data but selective causal stories that distort objective relationships by omitting relevant variables or misattributing causal directions. In their 2024 extension, the authors apply this logic to media markets, showing how media platforms strategically supply both biased information and empowering narratives to boost consumer engagement. The empirical study by \textcite{Andre2022} follows a similar logic. Specifically, the authors define narratives as backward-looking causal accounts of recent events, or explanations that people construct in order to make sense of what has happened, and that in turn shape their forward-looking expectations. The DAG-structure of narratives that is pervasive in this line of research can be viewed as a key contributor to the simplifying and organizing function narratives fulfill for individuals trying to make sense of information. \textcite{roos2024} synthesize these and other perspectives to propose a definition of collective economic narratives. They argue that economically relevant narratives are not just individual stories, but shared sense-making structures that arise in a social context, explain economic events, and suggest collective action. For a more detailed overview over different definitions of economic narratives, we refer readers to this work.

A recurring insight across the literature is that causality is the core ingredient of any narrative. It is what distinguishes a narrative from adjacent concepts such as topics or frames. While a topic identifies what is being discussed, a narrative links who did what to whom, and with what consequence. It attributes responsibility, defines trajectories, and frames problems in a way that invites action. Strikingly, \enquote{who is to blame for the problem} \parencite{crow2018narratives} is central to understanding how narratives influence public opinion and policymaking. Against this background, the present study puts particular emphasis on recognizing this specific property of narratives. In order to automatically identify narratives in large text data, detecting causal connections between events---implicit as well as explicit connections---is paramount.

Building on this literature, we define economic narratives as causal connections between two temporally and semantically distinct events, formulated in the structure \textit{A causes B} or \textit{A is caused by B}, that reflect an interpretative framing of economic developments. For example:
\begin{enumerate}
\itemsep -1em 
    \item the prices for cucumbers rose by 100\% this month - causes - people stop buying cucumbers
    \item increasing money supply - causes - prices for real estate go up
    \item the FOMC raised the policy rate - causes - turmoil on Wall Street

\end{enumerate}

Crucially, not every causal claim qualifies as a narrative in our framework. Narratives, as we understand them, are more than just descriptions of factually accurate chains of events; they often imply a perspective, or a selective emphasis that helps make sense of economic developments. To this end, for example, we distinguish between entities that qualify as genuine \textit{events} (such as \enquote{inflation is on the rise}) and those that do not (such as \enquote{high prices}, which has no temporal dimension). For a detailed operationalization of our definition, see \autoref{chap4}. %The goal is not to filter for only the most plausible or policy-relevant narratives, but to extract a wide range of causal claims as they appear in the text. In this sense, our approach prioritizes recall over precision, aiming to reflect the narrative density and diversity found in real-world discourse. Given this inclusive interpretation, we expect to find a substantial number of narratives in the dataset under analysis (see \autoref{chap3}).

\subsection{Extracting economic narratives from texts}

Historically, narrative extraction has been performed qualitatively. Hoping to create a more scalable solution, researchers have switched their focus to quantitative NLP methods to extract narratives from texts automatically. To be able to extract complex narratives however, classic NLP tasks such as topic modeling or sentiment analysis do not suffice. Each of these methods extracts information that is only a small part of most narrative definitions. For instance, topic modeling extracts latent topics from texts, enabling researchers to make an educated guess as to what the documents are about. This can be interpreted as the setting for a story, containing all necessary places and characters, but falls short of connecting these individual parts to a coherent sense-making story. 

As a result, researchers have increasingly turned to pipeline-based approaches. \textcite{ash21relatio} propose such a pipeline, named RELATIO, that heavily relies on semantic role labeling to extract narrative actors. They define narratives as a connection between so called \enquote{agents} and \enquote{patients}, where agents actively perform an action that affects patients. Notably, the causal component inherent in most theoretical approaches to narratives is not considered. For example, \textcite{ash21relatio} consider \enquote{ECB raise interest rate} to be a narrative, with \enquote{ECB} as the agent, \enquote{interest rate} as the patient and \enquote{raise} as the corresponding verb that defines the connection between the two. \textcite{lange22a} extend this concept by incorporating additional NLP tasks into the pipeline, such as coreference resolution and causal discovery, with the aim of connecting agent-patient pairs to form a sense-making story. However, they also acknowledge that complex pipelines with many interdependent components can cause even minor errors to cascade into serious downstream failures. This sensitivity has motivated the search for integrated models capable of handling these tasks within a unified framework.

Fixing this major issue of narrative extraction techniques, while simultaneously not watering down the definition, requires NLP models of a different architecture. We therefore propose to use LLMs to extract economic narratives from texts. For conceptual clarity, we use the term LLM only when referring to generative language models that have undergone instruction tuning. This limits the scope of the term to models like \texttt{GPT-4o}, while excluding families of discriminative language models like BERT. Given the \enquote{world knowledge} and language understanding capabilities encoded in LLMs, it is possible to prompt a model with the definition of an economic narrative and extract those narratives directly without the need to run the text through an error prone pipeline. Previous works have shown that LLMs are capable of solving related tasks, such as summarization, speaker attribution and Retrieval Augmented Generation \parencite{10.1145/3731445, fan2024surveyragmeetingllms, Bornheim_Grieger_Blaneck_Bialonski_2024}.

Recent research further underscores the potential of LLMs for content analysis. Studies in computational social science have shown that LLMs can match or even surpass human coders in annotating political, social, and economic texts \parencite{ziems2024can}. \textcite{mellon2024ais} reports that LLMs achieved 95\% agreement with expert annotators when analyzing British election statements. Similarly, \textcite{Gilardi_2023} demonstrates that LLMs like \texttt{GPT-3.5} could classify tweet content, author stances, and narrative frames more accurately than trained crowd workers. The applicability of LLMs to economic narratives is highlighted by \textcite{gueta2024can}, who use \texttt{GPT-3.5} to extract macroeconomic narratives from Twitter data. Although their operationalization of narratives relies heavily on sentiment analysis and topic modeling, their work demonstrates that LLMs are capable of handling unstructured textual data effectively. \textcite{schmidt2025narrating} makes use of the large language model (LLM) \texttt{Claude 3.5 Sonnet} to identify predefined inflation narratives in the German media coverage in 2022. Using a detailed codebook and exhaustive prompting techniques, the author was able to detect the same inflation narratives proposed by \textcite{Andre2022} and track their appearance in coverage, demonstrating that LLM prompting is a promising approach to extract backward-looking \enquote{blame} narratives from text.
\textcite{lange2025narrative} showed that combining scalable methods such as topic modeling with LLMs can lead to additional benefits and can be utilized to extract shifts in complex economic narratives over time. Instead of analyzing every available document with an LLM, the authors propose to only extract narratives at points in time in which a change in narrative is suspected. They detect change points in the word-topic distributions of the dynamic topic model RollingLDA \parencite{rieger22rolling} using bootstrap percentile tests \parencite{rieger22dynamic, lange22b}. Afterwards, the authors utilize the LLM \texttt{Llama 3.1 8B} \parencite{Llama3} to categorize the changes according to the Narrative Policy Framework from political science \parencite{jones2010narrative, shanahan2018narrative, schlaufer2022narrative}. Such an analysis can be adapted to work with other change detection methods \parencite[e.g.][]{benner22}, allowing researchers to use a detection method suited for their purposes. This approach of \textcite{lange2025narrative} does, however, also reveal the limitations of \texttt{Llama 3.1 8B}, as it infers narratives from most inputs, even when none are present. In our approach, which covers a more complex notion of an economic narrative, we therefore opt for a high-performing model to observe what kind of results a state of the art model can produce.

Given the increasing feasibility of narrative extraction through LLMs, a natural next step is to explore substantive domains where narratives matter most. The topic of inflation is particularly instructive in this regard, as few economic issues are more closely tied to public sentiment, expectation formation, and media framing.

\subsection{Inflation-related literature}

Inflation is particularly interesting in the context of narrative extraction. First and foremost, the topic is widely researched due to its formal relation to personal beliefs. In standard New Keynesian frameworks, expectations are not merely passive reflections of current conditions but actively shape future inflation trajectories \parencite{werning2022expectations}. Agents form beliefs about inflation that, in turn, influence their consumption, pricing, and wage-setting behavior \parencite{bachmann2015inflation, burch1975stock, juster1972inflation}. In this context, economic narratives play an important role, as narratives help individuals understand the world around them and make decisions. This tight linkage between belief formation and macroeconomic outcomes is one reason why inflation narratives have attracted more research attention than, say, labor market narratives. Central banks in particular have shown strong interest in the topic. In the modern forward guidance era, they aim to actively shape the inflation expectations of households and firms \parencite{blinder2008central}. Given this policy regime, it is unsurprising that a growing number of central bank researchers are engaged in studying the role of narratives in the expectation formation process \parencite{ter2022narrative, nyman2021news, kalamara2020making, tuckett2020monetary}.

One central finding from inflation expectation research is that media coverage appears to play an important role in expectation formation processes. As \textcite{conrad2022role} show in the German context, traditional media consumption is associated with more accurate perceptions of past and expected inflation, particularly when media coverage is intensive. Similarly, \textcite{lamla2014role} find that intensive inflation reporting increases the forecast accuracy of households, while negative or sensationalist framing can lead to exaggerated inflation perceptions. \textcite{ter2022narrative} further show that narrative monetary policy shocks (i.e., stories about interest rate decisions) can affect real macroeconomic variables if successfully disseminated.

Another central contribution in this domain is provided by \textcite{Andre2022}, who analyze the narratives individuals construct to explain the recent surge in inflation. Drawing on open-ended survey responses from thousands of U.S. households, the authors find that while expert narratives predominantly attribute inflation to demand-side factors (e.g., fiscal and monetary expansion), household and manager narratives are more heterogeneous and often emphasize supply-side shocks or political mismanagement. Through randomized priming experiments, the authors show that media narrative exposure causally shifts beliefs, highlighting the importance of inflation narratives in expectation formation.

Other lines of research suggests, however, that responsiveness of expectations to media coverage varies over time. Building on the rational inattention framework \parencite{reis2006inattentive, sims2003implications}, \textcite{coibion2015phillips} and \textcite{bracha2025inflation} provide evidence that attention to inflation is cyclical. As the authors show, media attention to inflation spikes in periods of high inflation and declines otherwise. Recent research by \textcite{schmidt2023inflation} references to the same effect by analyzing how varying levels of inflation reporting affect inflation expectations. The authors make use of the Inflation Perception Indicator (IPI), which tracks thematic shifts in German inflation coverage, to analyze narrative shifts in German inflation reporting.
%Based on the dynamic topic model RollingLDA \parencite{rieger22rolling}, the IPI captures both the intensity and thematic focus of inflation reporting over time---and allows for heuristic conclusions about the prevalent inflation related media narratives at each period in time.
Employing a threshold VAR framework, the authors show that the influence of media coverage on inflation expectations is regime-dependent: only during high-inflation periods does narrative intensity significantly affect expectations and real variables.

In what follows, we put these insights into practice. Using inflation as a well-studied and socially salient topic, we apply an LLM-based narrative extraction approach to a corpus of media reports. Although inflation serves as the empirical domain of this paper, our methodology is generalizable to other economic issues. Our goal is not only to assess the performance of LLMs in a high-stakes context, but also to demonstrate how economic narratives can be systematically identified in large text corpora.

\section{Model and data}
\label{chap3}

In our experiment, we aim to get the best performance possible given the current generation of LLMs. In this section, we argue why the model \texttt{GPT-4o} \citep{openai2024gpt4ocard} is our model of choice to accomplish this and describe how we have created the data set for our experiment.

\subsection{Choosing an LLM}
At the time of analysis, \texttt{GPT-4o} \citep{openai2024gpt4ocard} is the newest iteration in the 4th generation of the Generative Pre-trained Transformer family of models by OpenAI. We chose this particular LLM\textemdash model snapshot \texttt{gpt-4o-2024-11-20}  \textemdash over other commercial models, such as \texttt{GPT-4} \citep{openai2024gpt4technicalreport}, as it offers comparable or superior performance while being more scalable due to lower financial costs. Its performance in human reasoning and language understanding tasks is especially impressive \citep{GPT4oBenchmark}, as these are crucial components of narrative extraction, which requires deep understanding of complex economic circumstances. In particular, we choose to use \texttt{GPT-4o} over the more recent class of reasoning models, such as \texttt{OpenAI-o1} \parencite{o1}, as our prompt requires Chain-of-Thought-prompting \parencite{Wei2022} tailored specifically to our narrative definition. As recent research shows that prompting reasoning models with additional Chain-of-Thought sequences does not improve performance \parencite{wang2024advanced, deepseekai2025deepseekr1}, we opt for our own specialized prompting over the generic reasoning offered by models like \texttt{OpenAI-o1}. Our prompting strategy is described in detail in \autoref{chap5}.

While analyzing the potential of large commercial models to extract economic narratives is valuable, their use also entails notable downsides. For one, the cost of using such models on large data sets is very high and might not be feasible for universities, research centers or companies with low financial backing. Additionally, since these models are not publicly available and operate on proprietary hardware, their long-term availability is uncertain, as companies like OpenAI are likely to prioritize profit over maintaining access to older models. Therefore, all evaluations performed by older models will ultimately lose their reproducibility, undermining a cornerstone of good scientific practice. For this reason, the use of the commercial model \texttt{GPT-4o} in this paper should be understood as a theoretical example of what state-of-the-art LLMs are currently capable of, and what open-source models may achieve in the future as their development progresses.

\subsection{The text corpus}
We evaluate the proficiency of \texttt{GPT-4o} on the task of economic narrative extraction using a curated corpus of Wall Street Journal and New York Times news paper articles published between 01-01-1985 and 27-09-2023. We filtered for articles containing the words \enquote{inflation} or \enquote{price (increase|hike|surge)}. This time period was selected because it includes both high- and low-inflation phases. Focusing on inflation-related documents allows us to work with a thematically coherent corpus, enabling aggregation and comparison of narratives, even within a relatively small dataset.

To further narrow down the potential economic narratives in these documents, we provide the model with a short excerpt rather than the full article. Each excerpt includes the sentence containing the filter terms, along with the two preceding and two following sentences. This allows us to focus only on the topic of inflation while still providing enough context to interpret potential narratives, including those that span multiple sentences.

From this corpus, we randomly sample 100 documents to create gold-standard responses for the LLM. While a larger annotated dataset is always desirable, we prioritize a thorough annotation of a smaller subset over a broader but less precise coverage. The annotation task is complex, demands sustained focus, and is prone to errors when done hastily or without sufficient care.

\section{The codebook and the gold-standard}
\label{chap4}
When trying to operationalize rich concepts like economic narratives for empirical research, we often struggle to translate verbal definitions into a rigorous and workable codebook. Aspects of narratives that seem sensible enough in a verbal explanation are sometimes hard or even impossible to identify in an empirical setting. This problem gets exacerbated when a codebook is required to apply to LLM-based extraction in addition to human coders. While LLMs are trained to appear as human-like partners in interactions, they differ substantially from humans in the way they process information when performing tasks \parencite{Mondorf2024}.  

Therefore, we will proceed by clarifying what defining aspects of economic narratives we consider to be important in an empirical setting. Based on these aspects, we create a codebook designed for human coders to create a consensus gold-standard data-set. We then translate the human codebook into a prompt designed to maximize the LLM's performance, accounting for the model's peculiarities relative to human annotators.

\subsection{A narrative codebook}
Translating any of the narrative conceptualizations presented in \autoref{chap2} to empirical work entails some fundamental challenges. These definitions operate with high-level concepts such as \textit{event}, \textit{story} and \textit{moral}. As a result, when two experts are given only a definition to guide text annotation, their results are likely to differ significantly. Because a narrative is tied to language, domain understanding, and the reader's interpretation of the author's intent, no narrative extraction can be truly objective. Additionally, from a purely linguistic perspective, narrative extraction is not a simple span detection task in which one specific part of a text must be marked. Instead, key contextual information given in the text must be extracted and synthesized into a coherent structure. These structures must also be amenable to aggregation  when comparing narratives across large corpora.

We propose a codebook that both narrows down the type of narratives that are to be annotated and outlines a clear structure for the annotation process. The codebook is the end product of multiple workshops that were held with experts and non-experts. These sessions aimed to develop clear guidelines for extracting economic narratives, minimizing common coding errors and sources of ambiguity wherever possible. The entire codebook can be found in \autoref{CodeBook} in the appendix. The most important aspects of the codebook can be summarized as follows. 
\begin{itemize}
    \item \textbf{Goal}: The main task is to accurately extract narratives from newspaper excerpts, defined as a causal link between two consecutive events. Multiple researchers should code the same texts to minimize subjective bias.
    
    \item \textbf{Target Form}: Each narrative must be reduced to one of two forms: \textit{event 1 - causes - event 2} or \textit{event 1 - is caused by - event 2}. To preserve the natural order of the text, the first event in the source should remain as \textit{event 1} in the coded narrative.

    \item \textbf{Event Types}: Permissible events include occurrences, activities, conditions, future events, plans and policies. The coded event should remain as close as possible to the original wording, avoiding synonyms or abstractions.

    \item \textbf{Coreference Resolution}: When entities are referred to indirectly (e.g., with the use of pronouns), the coder is to replace them with the correct entity to maintain clarity (e.g., "\{He|President Biden\}"). 

    \item \textbf{Statements}: When an event consists of a statement made by an individual, coders must distinguish whether the causal link to another event pertains to the content of the statement or to the statement itself. Typically, the focus should be on the content. However, if the statement itself triggers a reaction, it should be coded as the event. Typical examples include forward-looking statements in corporate disclosures or forward guidance issued by central banks.

    \item \textbf{Embedded Clauses}: Coders should ignore non-essential elements within events (e.g., appositions, parentheses) unless they contribute meaningfully to the narrative. 
    
    \item \textbf{Chained narratives}: Each event in a chained narrative (e.g. \enquote{event 1 causes event 2; event 2 causes event 3}) should be coded separately.
    
    \item \textbf{Narrative forks}: When an event is caused by multiple events or an event causes multiple other events, each causal connection has to be noted as a separate narrative unless the combination of events is integral to the narrative.

    \item \textbf{Positive Causality Only}: Only positive causal links should be coded, negative ones (e.g., “does not cause”) should be excluded.

    \item \textbf{Economic Focus}: Only narratives with an explicit economic context should be coded. Non-economic narratives should be ignored.

    \item \textbf{Maintain Full Meaning}: The full message of the original narrative is to be retained without omitting crucial information, even if some parts seem superfluous.

    \item \textbf{Edge Cases}: When a code is unsure if a sentence contains a causal narrative, they code it anyway and resolve uncertainties later with other coders.
\end{itemize}

\subsection{Creating a gold-standard data set}

To annotate our 100 randomly sampled documents, three expert annotators \textemdash familiar with the definition of economics narratives, experienced in applying it and proficient in English at a near-native level\textemdash were provided with the codebook. Each annotator was tasked with coding all narratives from the 100 documents independently, without discussing their results with the other annotators. To address any remaining language barriers, such as idiomatic expressions or culturally specific references, annotators were given access to a built-in DeepL API, allowing them to translate the text into their native language if needed. In such cases, narratives were still coded based on the original English version.

In regular intervals during the annotation process, the annotators met to revise their understanding of the codebook and discuss edge cases they were uncertain about.

After coding all 100 documents individually, the annotators met in multiple sessions to discuss the results. Each individually coded narrative was discussed by all three annotators. If they unanimously agreed it was correctly coded, the narrative was added to the list of gold-standard narratives for the corresponding document. In this process, it did not matter whether a narrative had been identified by one, two, or all three annotators. Given the subjectivity of the task and the possibility that even experts may overlook narratives without considering multiple perspectives, inclusion of a narrative in the gold standard was based solely on substantive discussion grounded in the codebook. Such procedures have been shown to produce more valid and reliable results than simply applying majority voting \parencite{burla2008text, neidhardt2010selbststeuerung}.

The individual annotations by each annotator, however, were not discarded. Instead, we use them to compute an \textit{expected deviation} from the gold-standard, representing the level of variation that is acceptable among expert annotators. The output of the LLM is then compared to this baseline. We also document the reasons for each deviation from the gold-standard labels. Suppose an event has multiple consequences\textemdash i.e. \textit{Event A} causes three distinct events: \textit{B}, \textit{C} and \textit{D}. Annotator 1 did not split these into three individual narratives but instead coded \textit{Event A} as causing the combined outcome of \textit{B}, \textit{C} and \textit{D}. This would be classified as a \textit{minor deviation}, with the reason noted as \enquote{missing multiple consequences}. This approach of differentiating major from minor deviations allows us to formalize complex patterns within our codebook while still comparing human and AI performance effectively. \textit{Major deviations} either completely alter the meaning of a narrative, identify a narrative that does not exist or miss an existing narrative. Minor deviations, by contrast, occur when the overarching narrative is correctly identified but is either formally incorrect (e.g. due to missing coreference resolution) or contains subjective elements that differ from the gold-standard annotation.

In total, the gold-standard data set contains 291 narratives in 100 documents.

\section{Prompting LLMs to extract narratives}
\label{chap5}
Prompting language models to extract and annotate data in research is a relatively new venture. We feel strongly that LLMs open entirely new avenues for quantitative economic research. \textcite{Korinek2023} was among the first to urge economists to explore this potential. However, as of the time of this writing, the major use case for LLMs in economic research is sentiment analysis (Examples include \textcite{Jeong2025} and \textcite{Han2024}), for which high-quality models exist that do not require Generative AI. Due to the novelty of LLMs, no best practices for crafting prompts have been established to date in any of the social sciences. However, recent literature in AI and computational linguistics has introduced a set of general strategies that can enable researchers to make informed choices. We hope that our task can provide a useful example for related studies, and we will cite relevant work throughout. Using \texttt{GPT-4o}, we explore multiple approaches to extract economic narratives informally. We then narrow down the more useful principles which we subject to a formal evaluation. We require narratives to be extracted in a consistent format that can be parsed by machines. Therefore, we always use the \enquote{JSON} option in the OpenAI API, which guarantees that the model will output consistent Java Script Object Notation (JSON) files. 

\subsection{Prompt optimization: strategies and challenges}
\label{chap5.1}

The least data-intensive yet most heuristic prompting strategy is zero-shot learning \parencite{Wei2021, Kojima2022}. Generally speaking, LLM-assistants like ChatGPT are fine-tuned for instruction-following, which is the reason users can interact with them in a way that feels natural and human-like. By using zero-shot prompting, researchers can exploit this property by only supplying the model with the research question and task instruction along with the text to perform inference on. To set a benchmark for comparison to human performance, we use this zero-shot strategy, supplying our entire unabridged codebook as an instruction. The results are very inconsistent, both regarding the form of the outputs and in their faithfulness to the codebook.

We also explore a variation of this strategy, using a condensed version of our codebook. While current LLMs use interpolation techniques \parencite{Zhong2025} to expand hard context windows beyond the length of any common-sense codebook, it remains unclear how well performance is maintained as context size increases. Due to the intransitivity of LLM architectures, few theoretical guarantees can be considered for applied work. However, empirical investigations have documented phenomena such as the lost-in-the-middle effect \parencite{Liu2023a}, which suggests that increasing the length of inputs tends to result in subtle forms of information loss, mostly in the middle of input sequences. The distribution of information in prompts appears to be a relevant criterion as well. \textcite{Tian2024} suggest that a higher relative distance between crucial pieces of information in a prompt adversely affects information retrieval rates, which could compromise tasks like ours. Hence, ceteris paribus, limiting prompt length is advisable. However, the level of detail in our codebook must still cover the intricacies of the coding task. To find the sweet spot regarding this prompt-length trade-off, we condense our codebook for several iterations, emphasizing different aspects in each version. We evaluate by probing the outputs that result from using different codebook variations. 

Generally, we find that increasing the information density of the codebook tends to be beneficial for performance. However, a priori, it is not obvious what kinds of information should be compressed and what concepts must be dealt with extensively in the instructions. For example, several passages in the codebook deal with the definition of an event, including detailed elaborations on the types of events we consider to be the most relevant constituents of economic narratives. These include the implementation of a policy or the future-facing intentions of firms or policy makers. We endowed human coders with a great level of detail here partly because issues involving the delineation of events had come up regularly during the codebook workshops. However, as it turns out, \texttt{GPT-4o} defaults to a conceptualization of events that closely corresponds to our definition. Hence, simply mentioning that narratives consist of two events turns out to be the optimal level of information we can provide about events. By contrast, the ordering of events in the output, which we require to remain unchanged from the source text, turns out to be a problem that needs explicit mentioning in multiple spots of the instructions. 

In general, simply adding more requirements to prompts does not reliably increase LLM-performance. Additional aspects often introduce competing constraints that the model must resolve implicitly \parencite{yang2025promptsdontsayunderstanding}. Researchers should therefore prioritize which concepts to explain in detail. Morover, since an LLM's conceptual understanding is shaped by its training data and the alignment techniques used during training --- both of which are unknown to researchers --- practitioners should always probe the model's sensitivity to the way information is provided, even for seemingly uncontroversial concepts.

In addition to descriptive instructions, various sections of the codebook (see \autoref{CodeBook}) contain synthetic examples of full narratives and of narrative components. These examples are meant to highlight specific aspects of the extraction task for human coders to focus on when familiarizing themselves with the coding process. Such partial demonstrations are essential for designing high-quality codebooks, as they have shown to be helpful for human-coders \parencite{Saldana2016}. Conversely, they do not contribute positively to LLM performance on our task. The model appears to struggle with synthesizing these aspect-driven examples into a consistent, multi-aspect narrative definition, which adversely affects model performance.

\subsection{Few-shot Chain-of-Thought prompting}
Given the results of the initial explorations, we decide to fully adapt our original codebook for LLM inference. Instead of using instructions that mirror the structure of the original codebook outlined in \autoref{chap4}, we only provide the model with a succinct description of the key concepts and processes that it contains. Crucially, along with these instructions, we supply hand-labeled input-output examples. This widely used method is called few-shot learning in the AI literature (see \textcite{Song2023} for a review). Unlike the supervised learning paradigm that most empirical economists will be familiar with, effective few-shot learning only requires a handful\textemdash rather than hundreds\textemdash of hand-labeled training examples. In contrast to the strategy outlined in \autoref{chap5.1}, where distributed examples throughout the codebook each highlight a specific aspect, few-shots are complete input-to-output demonstrations that are provided in a dedicated block after the verbal instructions.

The few-shot paradigm allows the model to infer information about the desired output from full-fledged examples rather than from lengthy prosaic instructions. Consequently, the selection of few-shots and their total number are crucial. Few shots should be representative of the full data-set, ideally without adding redundancies. During this final round of prompt engineering, we use rigorous formal validation: From our set of 100 hand-annotated, gold-standard documents, we randomly select 20 for cross-validation. Any of these 20 examples can be used either as few-shots in prompts or for evaluating the performance of a candidate-prompt. The rest of the gold-standard documents are held out entirely for testing the performance of the final prompt. Given our validation set, we explore using between 1 and 9 few-shots in a single prompt. After evaluating the performance on the remaining validation example, holding constant the rest of the instructions, we settle for using 7 few-shots. Finally, we run evaluations with rotating sets of seven examples to identify the few-shot combination that yields the best evaluation results. Intuitively, the optimal set captures the most relevant narrative characteristics for the model to learn the task effectively.

We combine few-shot learning with the Chain-of-Thought (CoT) prompting paradigm, which means that we specify intermediate steps for the model to take when constructing an answer \parencite{Wei2022}, with the few-shot examples mirroring this step-wise format. When dealing with problems that require complex forms of reasoning, dividing up a task into sub-problems turns out to be advantageous. This is because LLMs can only \enquote{reason} about a problem insofar as they can produce tokens about it. The CoT-logic of transforming narrative extraction into a set of sequential tasks allows the model to devote more computational power to each step in the reasoning chain. Every step must be framed as a discrete action and output separately by the model before moving on to the next step. In addition to enhancing performance, CoT-prompting also increases transparency, as errors in LLM inference can be traced to particular steps in the chain. 

Conceptually, the CoT paradigm is reminiscent of classic pipelines approaches to economic narrative extraction (see \autoref{chap2}), where sub-tasks are handled by different language models rather than a unified LLM \parencite{ash21relatio, lange22a}. However, the CoT pipeline merely imitates a full separation into sub-task. The LLM always has access to the prior steps since they are stored in its context. Since our codebook contains interdependent tasks that require economic judgment and adherence to linguistic principles, this feature is highly desirable. 

\subsection{An integrated narrative extraction pipeline}
Our final prompt relies heavily on the few-shot CoT paradigm introduced in the previous section. A detailed description can be found in \autoref{Prompt} in the appendix. The prompt is divided into two main parts, \textit{Codebook} and \textit{Examples}, where the former subsumes the written instructions and the latter only encompasses the few-shots. 

The first subsection of the codebook, \textit{Basic Idea}, is designed to prime the model by broadly outlining the task and the source medium of the input texts. Subsection 2, \textit{Definition}, briefly states the crucial structural elements needed in every extracted narrative:

\begin{center}\textit{An economic narrative consists of exactly two events \\ and a causal connection that is asserted between those events.}\end{center}

Subsection 3 is called \textit{Event structures}. It introduces three distinct types of linguistic constructions in which causally linked events are typically embedded when they are discussed within the article excerpts. We build on the formalism of graphical models introduced by \textcite{Pearl2009}, which is widely used in the AI field and in parts of economics, to distinguish \textit{direct causal effects} from causal \textit{chains} and \textit{forks}. A causal chain occurs if an \textit{Event A} causes an \textit{Event B} which in turn causes an \textit{Event C}. A causal fork occurs if an \textit{Event B} is a common cause for two Events, \textit{A} and \textit{C}. Embracing this terminology enables us to represent causal structures as directed acyclic graphs (DAGs), as is common practice in the literature on economic narratives. Using this established formal language also allows us to optimally leverage \texttt{GPT4o's} pre-training, because we do not have to explain concepts in the instructions to the LLM as long as they are encapsulated in the model's world knowledge. As stated in \autoref{chap4}, we require chains and forks to be coded as separate narratives by the model. While some of the few-shots contain examples of such causal constellations, the model turned out to struggle with them, which prompted us to incorporate them into the \textit{Codebook} section. Hence, apart from raising awareness about the aforementioned causal construct, subsection 3 also sets up expectations on how the LLM should deal with these structures when they arise.

Subsections 4 and 5, \textit{Rules} and \textit{Target forms}, precisely establish the set of allowable forms that extracted narratives are expected to conform to. We assert that the expression denoting the causal connection between events in a narrative must be classified into either \enquote{causes} or \enquote{caused by}, depending on the direction of the causal relationship. As laid out in \autoref{chap2}, causality is a core ingredient of economic narratives, because causal connections enable individuals to make sense of why events have unfolded. Many economic choices by households and firms are based on this sense-making inference. Quite commonly, however, causal connections that appear obvious to human readers are not written down explicitly in the source text. One of the main advantage of using LLMs for our task is their ability to recognize a diverse set of explicit and implicit causal connections from a given context. While extracting implicit causality\textemdash when related events are not connected using a specific causal cue\textemdash is challenging, it is equally important. For example, when events occur at different points in time, authors often imply causal relationships through temporal cues. Distinguishing whether the author intends to signal causality or merely describes a sequence of independent events often depends on contextual knowledge. LLMs excel at such tasks, which require deep language understanding and nuanced interpretation. This capability sets them apart from dictionary-based methods that rely on explicit cue words to detect causality or language models like BERT \parencite{Bert} that require fine-tuning.

\begin{figure}[t]
  \centering
  % Chain-of-Thought Logic Diagram
% Author: GPT
% Compile with: pdflatex or lualatex
\definecolor{mint}{HTML}{72C2A4}
\definecolor{magenta}{HTML}{C27290}
\begin{tikzpicture}[
  token/.style={text=gray!170, minimum height=14pt, inner sep=8pt, font=\footnotesize, align=center},
  input/.style={draw, minimum height=14pt, inner sep=5pt, font=\footnotesize, align=center},
  steps/.style={draw, text=gray!170, rounded corners=2pt, inner sep=3pt, font=\scriptsize, align=center},
  tag/.style={draw, fill=gray!70, text=white, rounded corners=2pt, inner sep=4pt, font=\footnotesize, align=center}
]
  % ---------------------- Top row: high‑level CoT steps ----------------------
  \node[token] (s1) {Focused\\Excerpt};
  \node[token, below=28pt of s1] (s2) {Sequence\\of Interest};
  \node[token, below=28pt of s2] (s3) {Causal\\Restatement};
  \node[token, below=28pt of s3] (s4) {Coreference\\Resolution};
  \node[token, below=28pt of s4] (s5) {Event\\Rephrasing};

  % Step labels under each box (mimicking SRL argument tags)
  \node[steps, above=-3pt of s1] {Step 1};
  \node[steps, above=-3pt of s2] {Step 2};
  \node[steps, above=-3pt of s3] {Step 3};
  \node[steps, above=-3pt of s4] {Step 4};
  \node[steps, above=-3pt of s5] {Step 5};

  % ---------------------- Detailed representation for Step 3 ----------------------
  \node[input, right=34pt of s1, yshift=10pt] (FE) {The Fed has finally committed to a more stringent path.\\ \colorbox{mint}{It has raised short-term rates, hoping to ease inflationary pressures}\\that have been fueled by increasing oil prices.};
  
  \node[input, right=28pt of s2, yshift=10pt] (SoI) {\textit{It has raised short-term rates, \colorbox{mint}{hoping to ease} inflationary pressures.}};

  \node[input, right=0.6cm of s3, yshift=10pt] (CR_A) {\colorbox{mint}{It} has raised short-term rates};
  \node[input, right=0.3cm of CR_A] (CR_conn) {\textit{causes}};
  \node[input, right=0.3cm of CR_conn] (CR_B) {ease inflationary pressures};

  \node[input, right=0.6cm of s4, yshift=10pt] (CoR_A) {\textit{The Fed} \colorbox{mint}{has raised} short-term rates};
  \node[input, right=0.3cm of CoR_A] (CoR_conn) {causes};
  \node[input, right=0.3cm of CoR_conn] (CoR_B) {\colorbox{mint}{ease} inflationary pressures};

  \node[input, right=0.6cm of s5, yshift=10pt] (ER_A) {The Fed \textit{raising} short-term rates};
  \node[input, right=0.3cm of ER_A] (ER_conn) {causes};
  \node[input, right=0.3cm of ER_conn] (ER_B) {\textit{easing} inflationary pressures};

  \node[tag, below=-1pt of CR_A] {Event A};
  \node[tag, below=-1pt of CR_conn] {Connector};
  \node[tag, below=-1pt of CR_B] {Event B};

  \node[tag, below=-1pt of CoR_A] {Event A};
  \node[tag, below=-1pt of CoR_conn] {Connector};
  \node[tag, below=-1pt of CoR_B] {Event B};

  \node[tag, below=-1pt of ER_A] {Event A};
  \node[tag, below=-1pt of ER_conn] {Connector};
  \node[tag, below=-1pt of ER_B] {Event B};

  % ---------------------- Arrows indicating process flow ----------------------
  \draw[-{Stealth[length = 6pt, width = 7pt]}]
      ($ (FE.south west) + ( 45pt,15pt ) $)  % 45 pt right, 4 pt down from FE.south-west
  --  ($ (SoI.north west) + ( 65pt,0pt ) $); % 65 pt right, 4 pt up   from SoI.north-west
  \draw[-{Stealth[length = 6pt, width = 7pt]}]
      ($ (SoI.south) + ( 22pt,6pt ) $) 
  --  ($ (CR_conn.north) + ( 0pt,0pt ) $);
  \draw[-{Stealth[length = 6pt, width = 7pt]}]
      ($ (CR_A.south west) + ( 15pt,6pt ) $) 
  --  ($ (CoR_A.north west) + ( 26pt,-7pt ) $);
  \draw[-{Stealth[length = 6pt, width = 7pt]}]
      ($ (CoR_A.south west) + ( 60pt,6pt ) $) 
  --  ($ (ER_A.north west) + ( 60pt,-6pt ) $);
  \draw[-{Stealth[length = 6pt, width = 7pt]}]
      ($ (CoR_B.south west) + ( 18pt,6pt ) $) 
  --  ($ (ER_B.north west) + ( 28pt,-6pt ) $);

\begin{scope}[on background layer]
  \node[fill=gray!8, rounded corners=4pt, inner sep=6pt,
        fit=(s1) (FE) (s5) (CoR_B)] {};
\end{scope}

\end{tikzpicture}
  \caption{A diagram depicting the Chain-of-Thought transformations that our prompting strategy induces. The chain extracts an economic narrative from an example document and molds it into a standardized form step by step. Green highlighting indicates segments to be changed in the subsequent step, arrows map the source segments to their corresponding results. Results are cursive.}
  \label{fig:cot}
  \vspace{1em}
\end{figure}
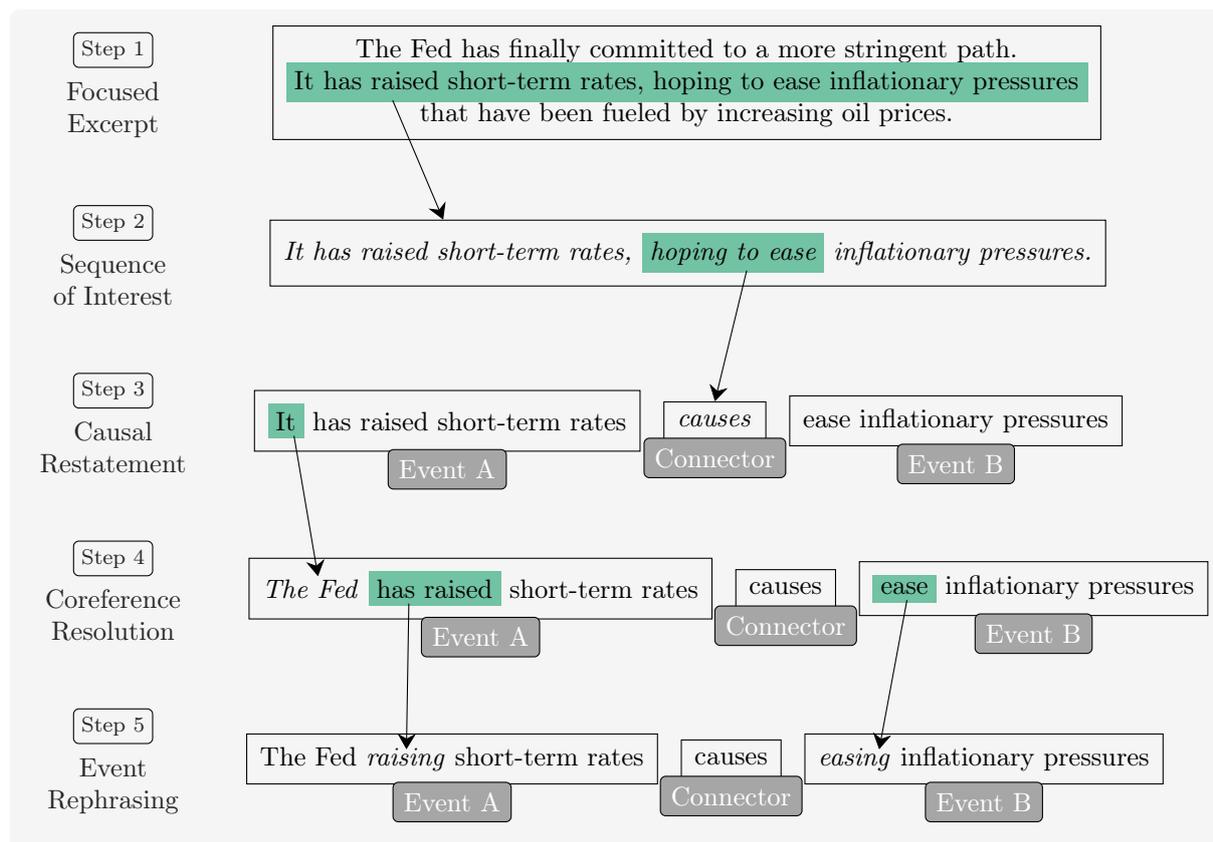

In the final subsection, \textit{Extraction process}, we describe the CoT that the model should produce for every input. \autoref{fig:cot} displays this procedure for an example text. 

As mentioned before, LLM performance hinges on steering the model's attention mechanism to relevant sections of the inputs. Step 1, therefore, consists of extracting a \textit{focused excerpt} from the input, which is an article fragment that features a term deemed indicative for an inflation discourse. Since the input texts vary significantly in their narrative density, it is helpful to first let the model disregard any sections from it that do not contain any events or causal relationship. This initial recognition of a narrative sequence is already non-trivial. In our example, the events are the Fed's realized policy measure\textemdash raising interest rates\textemdash and it's intention of easing inflationary pressures in the economy. Their causal connection is hypothetical because the outcome event has not occurred and is not guaranteed to occur at all. These nuances must be recognized by the LLM in the first CoT step, at least to the degree that it will include the segment in the focused excerpt instead of disregarding it altogether. 

Step 2 delineates a single, previously unextracted narrative from the focused excerpt by producing a \textit{Sequence of interest}. The task requires the LLM to isolate exactly two events that are causally connected, while disregarding any unrelated elements found in the text. Sequences of interest are essentially full narratives in their rawest form. In our example, the focused excerpt actually features two more events\textemdash the Fed committing to a more stringent path and increasing oil prices. While either of those might be important for further narratives, Step 2 requires the LLM to not attend to them for the time being. The \textit{Causal restatement} in Step 3 serves to separate the events from the causal connector and to make the latter explicit. Collectively, these first 3 CoT steps perform the \enquote{heavy lifting} of narrative recognition. 

Steps 4 and 5 only apply for a subset of extracted narratives. Steps 4 replaces pronouns in the sequence of interest with the entities they refer to. This task is known as \textit{coreference resolution} in the NLP literature and is generally handled with high accuracy by the LLM. If required, this step is crucial because entities that act or are acted upon are essential component of many narratives \parencite{Gehring2023}. Step 5 consists of some final grammatical adaptations: For the sake of event aggregation, we require events to be grammatically interchangeable. \textit{Event rephrasing} standardizes grammatical structure, allowing for easier extraction, comparison, and inference over events. This step mostly entails nominalization, in which a verb phrase is rephrased as a noun phrase.

After Step 5, we redirect the LLM to the focused excerpt to check for further narratives.

To validate our prompting strategy, we use between 1 and 9 few-shots from the validation sample and evaluate performance on the remaining 11 examples. In our experiments, performance peaks with 7 few-shots. Subsequently, we test and evaluate \texttt{GPT-4o} with our chosen prompt on the remaining 80 documents with gold-standard labels. We set the LLMs temperature hyperparameter, which governs the randomness or creativity of its responses, to 0.2. This setting makes responses more deterministic than the default while retaining enough variability to allow the model to explore the space of possible answers. For the purpose of reproducibility, eliminating any randomness by using a temperature setting of 0 would be preferable. However, a fully deterministic setting is known to degrade performance of generative language models, causing what is known as the \textit{likelihood trap} \parencite{Huang2023a, zhang2021}.

% \section{Evaluation}
% \label{chap5}

\subsection{Narrative abstraction and event clustering}
\label{chap4:abstraction}

After extracting all narratives in the form of \textit{Event A causes Event B} from the 80 test documents, we apply a series of post-processing steps aimed at enabling large-scale abstraction and aggregation of recurring economic patterns. The goal of this procedure is to move beyond individual instances and identify broader narrative structures. The following steps represent an initial attempt to structure and cluster narrative content in a principled way. While the primary focus of this paper lies in the extraction of narrative pairs, we briefly outline this pipeline to illustrate how the extracted data can be further processed and organized for downstream applications.

\textbf{Event decomposition.}  
First, we split compound events into distinct atomic propositions. Events such as \enquote{Energy and food prices are on the rise} were forked into two independent entries, such as \enquote{Energy prices are on the rise} and \enquote{food prices are on the rise}. This decomposition is essential, as early tests revealed that our narrative extraction model occasionally fails to fork such multi-entity constructions in the desired way (see above). Accordingly, this step functions as a correction layer that may become redundant in future implementations if the model's extraction capabilities improve. We implement this step using a few-shot prompt with \texttt{GPT-4o-mini}, designed to detect conjunctions and rewrite them into separate, grammatically coherent event statements. This approach ensures that each entry represents a clearly interpretable unit of analysis. By contrast, a simple regex-based \enquote{and}-splitting approach would often yield semantically incomplete fragments, such as \enquote{Energy} in the upper example, that lack narrative coherence.

\textbf{Valence and topic assignment.}  
Next, each atomic event is passed through another LLM-based classification step to extract both its semantic \textit{topic} (e.g., \enquote{inflation}, \enquote{interest rates}) and its directional \textit{valence} (e.g., \enquote{rising}, \enquote{falling}, \enquote{high}). The prompt follows a structured format with few-shot examples and requires the model to output a JSON object containing both events A and B and their respective valences, thereby ensuring consistent outputs across all extractions.

\begin{figure}[ht]
  \vspace{1em}
  \centering
  \usetikzlibrary{positioning,fit}

\begin{tikzpicture}[
  cluster/.style={
    draw,
    thick,
    rounded corners,
    align=left,
    font=\small,
    inner sep=4pt,
    minimum width=4.5cm,
    node distance=0.4cm and 0.4cm,
  }
]

% Bond Market Cluster
\node[cluster] (bond) {
  \textbf{Cluster: Bond Market}\\[2pt]
  • bond market\\
  • bond markets\\
  • Treasury bonds\\
  • ...
};

% Consumer Spending Cluster
\node[cluster, right=of bond] (cons) {
  \textbf{Cluster: Consumer Spending}\\[2pt]
  • consumer spending\\
  • consumer demand\\
  • household spending\\
  • ...
};

% Energy Prices Cluster
\node[cluster, right=of cons] (energy) {
  \textbf{Cluster: Energy Prices}\\[2pt]
  • energy prices\\
  • electricity prices\\
  • home energy prices\\
  • wholesale energy prices\\
  • fuel prices\\
  • ...
};

\end{tikzpicture}
  \caption{Example illustrating how events are grouped by a cluster topic.}
  \label{fig:clustering}
  \vspace{1em}
\end{figure}
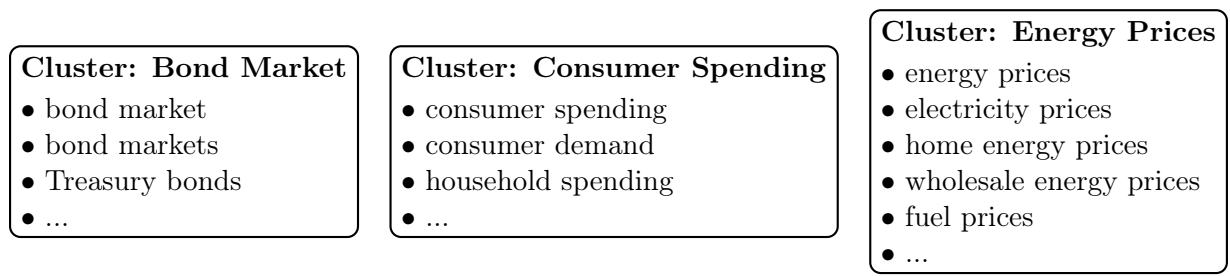

\textbf{Topic normalization and abstraction.}  
Since we prompted our initial narrative extraction model to stick as closely to the original wording of the text as possible during inference, many of the extracted narratives vary lexically while describing the same or similar phenomena (e.g., \enquote{interest rates} vs. \enquote{borrowing costs}). Therefore, we implement a semi-automated abstraction step to cluster semantically similar topics. First, we embed all extracted topic strings using the \texttt{all-MiniLM-L6-v2} sentence-transformer model. Second, we map each topic embedding to a controlled set of embeddings representing different macroeconomic categories (e.g., \enquote{government spending}, \enquote{interest rates}, \enquote{inflation}). To compare the topic embeddings and the predefined anchor terms, we use cosine similarity metrics, as they are invariant to the high number of embedding dimensions. Previous research has shown that clustering by anchor terms can serve as an alternative to fine-tuning based classification approaches in an unsupervised setting and help to interpret embeddings \citep{lex2sent, POLAR2}. We take care to preserve economically meaningful distinctions, e.g., \enquote{inflation expectations} and \enquote{inflation} remain separate due to their distinct roles in economic theory and policy, and even in public understanding. We then manually refine the mappings by carefully inspecting cluster compositions. In cases where embedding similarity alone appears insufficient, we use domain knowledge and manually add recurrent topic synonyms to the respective topic cluster or remove misclassified terms. Via pattern-based string matching we then replace all topics that were assigned to a certain cluster with the respective topic label. This hybrid approach ensures that topic abstraction remain both semantically robust and economically meaningful. An example of this mapping is provided in \autoref{fig:clustering}.

Finally, we translate all valence expressions indicating an increase or general high level of something with an upwards-pointing arrow (↑) and all expressions indicating low levels or a decrease with a downwards pointing arrow (↓). In cases where implicit negations are involved, these are reversed to ensure that, for example, \textit{rising economic vulnerability} and \textit{rising economic stability} are represented by different arrows, even if the term \enquote{rising} is used in both contexts.

\begin{figure}[t]
  \centering
  % ------------------------------------------------------------------
% Narrative-Abstraktion (Beispiel Art. 1664341137_21213)
% ------------------------------------------------------------------
\definecolor{mint}{HTML}{72C2A4}      % Grün
\definecolor{magenta}{HTML}{C27290}   % Magenta
\definecolor{steel}{HTML}{556F7A}     % dezentes Grau-Blau

\begin{tikzpicture}[
  box/.style   ={draw, rounded corners=3pt, inner sep=6pt,
                 font=\footnotesize, align=center, text width=12cm},
  tag/.style   ={draw, fill=steel, text=white, rounded corners=2pt,
                 inner sep=4pt, font=\scriptsize},
  step/.style  ={font=\scriptsize\bfseries, text=gray!60},
  arrow/.style ={-{Stealth[length=6pt,width=7pt]}, very thick, color=steel},
  token/.style={text=gray!170, minimum height=14pt, inner sep=8pt,
        font=\footnotesize, align=center},
  steps/.style={draw, text=gray!170, rounded corners=2pt, inner sep=3pt,
        font=\scriptsize, align=center},
]

% ------------------ Zeilen-Layout ------------------
\node[box] (raw)  {\textbf{Excerpt}\\[3pt]
  \emph{``Some private economists -- and a few inside the Fed -- say}
  \colorbox{mint}{the Fed's aggressiveness} \emph{is increasing}
  \colorbox{magenta}{the risks of an outbreak of inflation}.''};

\node[box, below=5mm of raw] (events) {%
  \textbf{Extracted Narratives}\\
  \texttt{7-shot CoT based GPT-4o}\\[4pt]
  Event A: \colorbox{mint}{the Federal Reserve's aggressiveness}\\
  Event B: \colorbox{magenta}{higher risks of an outbreak of inflation}};

\node[box, below=5mm of events] (labels) {%
  \textbf{Topic-Valence Pairs}\\
  \texttt{Few-shot GPT-4o-mini}\\[4pt]
  \begin{tabular}{@{}l l@{}}
    \textcolor{mint}{loose monetary policy}
    \hspace{2mm}causes\hspace{2mm}
    \textcolor{magenta}{rising inflation risks}
  \end{tabular}};

\node[box, below=5mm of labels, minimum height=20pt] (norm) {%
  \textbf{Normalized Narrative}\\
  \texttt{Embedding sim.\& dictionary}\\[4pt]
  \textcolor{steel}{\Large\(\uparrow\)}\,\textbf{monetary policy}
  \(\;\longrightarrow\;\)
  \textcolor{steel}{\Large\(\uparrow\)}\,\textbf{inflation}};

% ------------------ Schritt-Labels ------------------
% \node[step, left=3mm of raw.west] (s1) {Step 1};
% \node[step, left=3mm of events.west] (s2) {Step 2};
% \node[step, left=3mm of labels.west] (s3) {Step 3};
% \node[step, left=3mm of norm.west] (s4) {Step 4};

\node[token, left=5mm of raw.west, yshift=-3mm] (s1) {Original\\Statement};
\node[token, left=5mm of events.west, yshift=-3mm] (s2) {Narrative\\Extraction};
\node[token, left=4mm of labels.west, yshift=-3mm] (s3) {Post\\processing I};
\node[token, left=3mm of norm.west, yshift=-3mm] (s4) {Post\\processing II};

% Step labels under each box (mimicking SRL argument tags)
\node[steps, above=-3pt of s1] {Step 1};
\node[steps, above=-3pt of s2] {Step 2};
\node[steps, above=-3pt of s3] {Step 3};
\node[steps, above=-3pt of s4] {Step 4};

% ------------------ Verbindungs-Pfeile ------------------
\draw[arrow] (raw.south)    -- (events.north);
\draw[arrow] (events.south) -- (labels.north);
\draw[arrow] (labels.south) -- (norm.north);

% ------------------ Hintergrundrahmen ------------------
\begin{pgfonlayer}{background}
  \node[fill=gray!6, rounded corners=5pt, inner sep=6pt,
        fit=(s4) (raw) (norm) (norm)] {};
\end{pgfonlayer}

\end{tikzpicture}
  \caption{Illustration of our narrative abstraction pipeline using a real example from the corpus.}
  \label{fig:transit}
  \vspace{1em}
\end{figure}
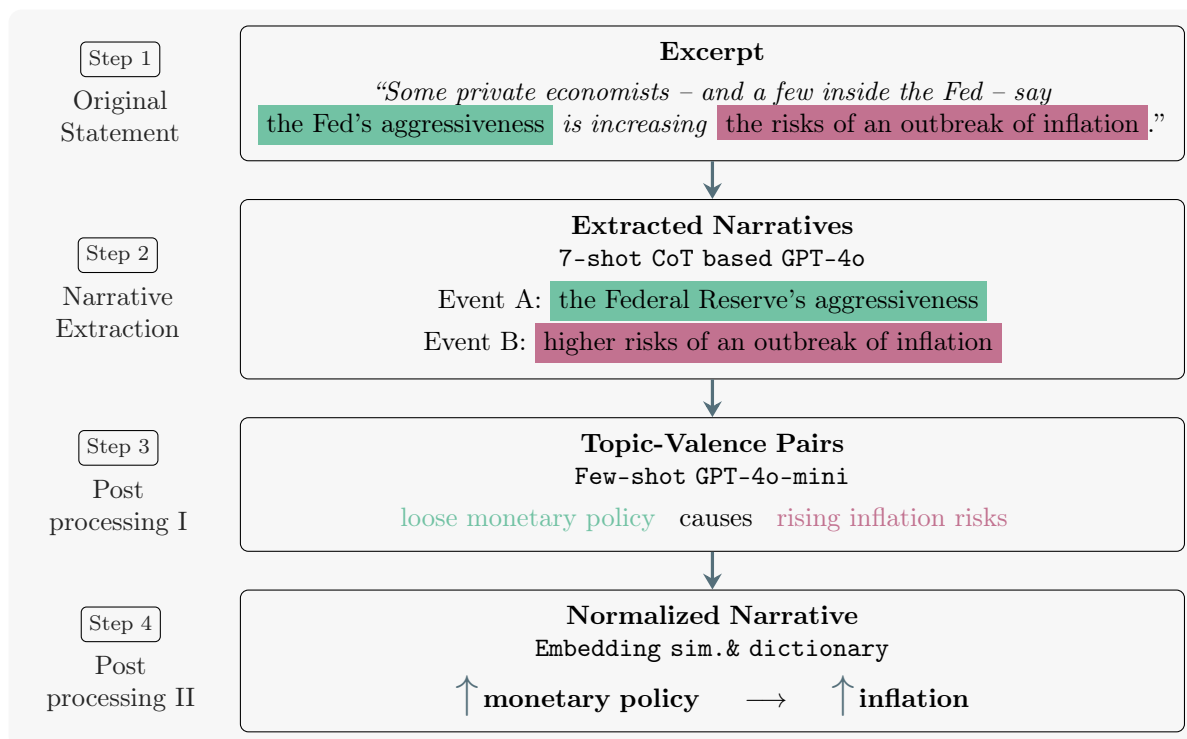

We are well aware of the information loss this step inevitably entails and the economic limitations that come with it. For instance, real interest rates differ fundamentally from nominal interest rates, which in turn are not the same as borrowing costs. However, we argue that such simplifications are justified given the inherent nature of narrative formation (i.e., the tendency to reduce complexity) and the context of our corpus, which consists of general-interest newspaper articles. Distinctions that are highly relevant in academic economics (e.g., between monetary aggregates or types of interest rates) may be less salient or even irrelevant to the typical audience of U.S. dailies.
In some cases, however, this loss of specificity results in tautological or arbitrary narrative formulations, such as \textit{rising interest rates cause higher interest rates}. While such artifacts are rare in our dataset, they illustrate one of the limitations of semantic abstraction: when the model reduces different expressions to a shared label, it may inadvertently collapse distinct concepts into self-referential loops. We acknowledge this as a trade-off inherent to our current aggregation strategy, though it affects only a small number of cases and might be mitigated through more refined post-processing in future applications.
Figure~\ref{fig:transit} visualizes the final extraction pipeline using a real example from our corpus.

\section{Results}
\label{chap6}
Using the codebook presented earlier, we let \texttt{GPT-4o} process the remaining 80 documents that were neither used for evaluation nor for model prompting. The results offer a nuanced picture of the model's ability to extract economic narratives from text. We will first discuss the results that we have attained through prompting the LLM and then turn to the critical post-processing step of aggregating the extracted narratives. Despite the complexity involved in creating the gold-standard, comparing \texttt{GPT-4o} to this benchmark turned out to be straight-forward, since essentially all disagreements had been resolved beforehand.

The dataset containing test and evaluation documents, including the gold standard as well as the model's and individual experts' annotations, is publicly available and can be found on \href{https://doi.org/10.5281/zenodo.15622517}{GitHub}.

\subsection{Narrative identification}

On a structural level, the model performs remarkably well. In almost all cases, \texttt{GPT-4o} precisely follows the formatting instructions and consistently produces valid JSON outputs. In \autoref{AnnotatedExample} of the appendix, we present three examples comparing the gold standard narratives to those generated by the model, alongside the original excerpt for reference. 

The LLM also closely adheres to the CoT-procedure laid out in \autoref{fig:cot}. In Steps 1 and 2, \texttt{GPT-4o} typically paraphrases relevant parts of the source text in a semantically faithful and economically coherent manner. The model also correctly reproduces event order and directions of causality in most narratives. The CoT logic likely aids this process, as it requires the model to state the narrative sequence as continuous text twice before imposing causal structure and determining event order in Step 3. Even when the model misorders events, it typically preserves correct causal directionality, making such errors relatively minor and more grammatical than substantive. Misorderings tend to favor the causal connector \textit{causes} over \textit{caused by}. The reason for this preference is speculative. \texttt{GPT-4o} may favor stating causes first and effects second because such constructions occur more frequently in its training data.

In \autoref{tab:summary_stats}, we compare the model's performance to that of our expert annotators' using basic summary statistics. Notably, the LLM identifies a similar number of narratives per document as the human coders, on average. This suggests that the overall level of performance is not compromised by persistent over- or under-identification of narratives. However, a closer look at the distribution reveals important limitations. While human annotators show substantial variance in how many narratives they identify per document (standard deviation of 2.03, on average) the model's output is remarkably stable, averaging 2.32 narratives per document with a standard deviation of only 1.15. This points to a potential inductive bias: the model appears to have an implicit prior about how many narratives it \enquote{expects} to find, leading to systematic over-identification in narrative-sparse texts and under-identification in narrative-dense ones. This pattern is also evident in edge cases. In four documents where the gold standard specifies no narrative, \texttt{GPT-4o} nonetheless extracts at least one narrative in each case. This suggests that the model may be prone to overfitting the 7-shot examples provided in the prompt. 

\begin{table}[t]
\centering
\begin{tabular}{lccccc}
\toprule
\textbf{Measure} & \textbf{Expert 1} & \textbf{Expert 2} & \textbf{Expert 3} & \textbf{Gold} & \textbf{Model} \\
\midrule
Average  & 2.22 & 2.36 & 2.29 & 2.91 & 2.32 \\
Standard deviation           & 1.95 & 2.12 & 2.04 & 2.35 & 1.15 \\\bottomrule
\end{tabular}
\vspace{0.5em}
\caption{Narrative counts per document for human experts and \texttt{GPT-4o}.}
\label{tab:summary_stats}
\end{table}

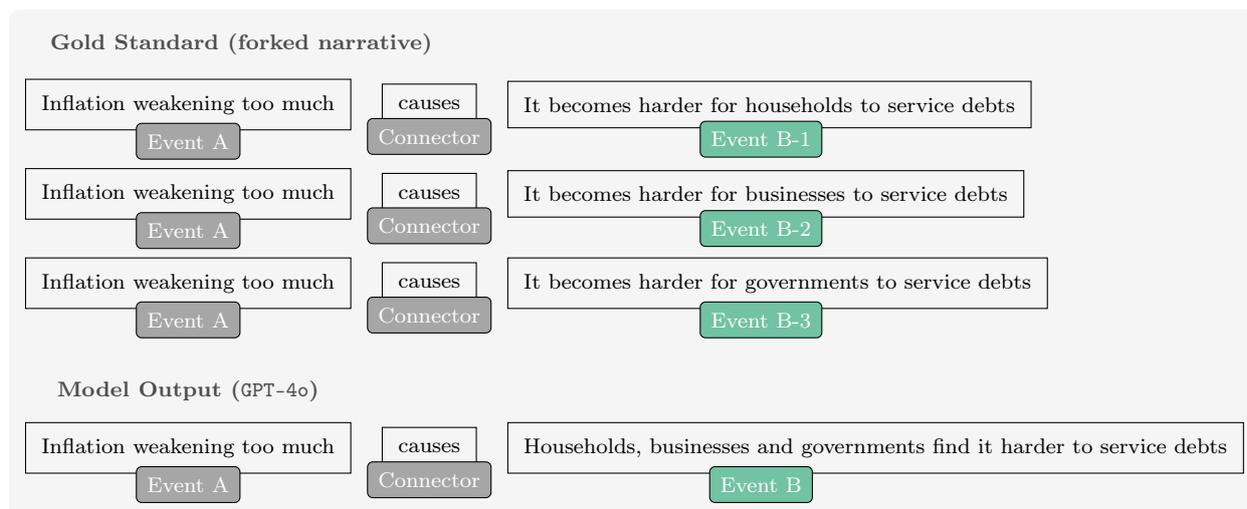
\begin{figure}[ht]
  \centering
\vspace{1em}
  \definecolor{mint}{HTML}{72C2A4}

\scriptsize
\begin{tikzpicture}[
  input/.style={draw, minimum height=14pt, inner sep=6pt, font=\scriptsize, align=center},
  tag/.style={draw, fill=gray!70, text=white, rounded corners=2pt, inner sep=4pt, font=\scriptsize, align=center},
  tagmint/.style={draw, fill=mint, text=white, rounded corners=2pt, inner sep=4pt, font=\scriptsize, align=center},
  label/.style={font=\scriptsize\bfseries, text=gray!140}
]
% --- Titel
\node[label] at (0.7, 6.8) (g_title) {Gold Standard (forked narrative)};

% --- Gold Standard (3 Zeilen)
\node[input] (g1a) at (0,6) {Inflation weakening too much};
\node[input, right=0.4 of g1a] (g1b) {causes};
\node[input, right=0.4 of g1b] (g1c) {It becomes harder for households to service debts};

\node[input, below=0.5 of g1a] (g2a) {Inflation weakening too much};
\node[input, right=0.4 of g2a] (g2b) {causes};
\node[input, right=0.4 of g2b] (g2c) {It becomes harder for businesses to service debts};

\node[input, below=0.5 of g2a] (g3a) {Inflation weakening too much};
\node[input, right=0.4 of g3a] (g3b) {causes};
\node[input, right=0.4 of g3b] (g3c) {It becomes harder for governments to service debts};

% --- Tags Gold
\node[tag, below=-0.1 of g1a] {Event A};
\node[tag, below=-0.1 of g1b] {Connector};
\node[tagmint, below=-0.1 of g1c.south west, xshift=95pt] {Event B-1};

\node[tag, below=-0.1 of g2a] {Event A};
\node[tag, below=-0.1 of g2b] {Connector};
\node[tagmint, below=-0.1 of g2c.south west, xshift=95pt] {Event B-2};

\node[tag, below=-0.1 of g3a] {Event A};
\node[tag, below=-0.1 of g3b] {Connector};
\node[tagmint, below=-0.1 of g3c.south west, xshift=95pt] {Event B-3};

% --- Titel Model
\node[label] at (0, 2.2) (m_title) {Model Output (\texttt{GPT-4o})};

% --- Chatty Output (eine Zeile)
\node[input, below=1.5 of g3a] (m1a) {Inflation weakening too much};
\node[input, right=0.4 of m1a] (m1b) {causes};
\node[input, right=0.4 of m1b] (m1c) {Households, businesses and governments find it harder to service debts};

% --- Tags Model
\node[tag, below=-0.1 of m1a] (eva) {Event A};
\node[tag, below=-0.1 of m1b] {Connector};
\node[tagmint, below=-0.1 of m1c.south west, xshift=95pt] (evb) {Event B};

\begin{scope}[on background layer]
  \node[fill=gray!8, rounded corners=4pt, inner sep=6pt,
        fit=(g1a) (g_title) (eva) (m1c)] {};
\end{scope}

\end{tikzpicture}
\normalsize
  \caption{Comparison of a forked narrative (as indicated by the gold standard) and a non-forked narrative (as returned by the LLM).}
  \label{fig:forked_example}
  \vspace{1em}
\end{figure}

To better understand this pattern, it is useful to examine some specific challenges the model faces. The lack of variability in the number of narratives is partly due to the model struggling with causal chains and forks. Even with explicit instructions and a dedicated CoT step in place, the model tends to return most narratives in a non-forked fashion, even in clear-cut cases where all human experts agree. Rather than disentangling forked or chained narratives into multiple atomic phrases, the model frequently compresses them into a single generic statement (see \autoref{fig:forked_example}).
This issue is related to stylistic differences across documents. Some of the articles break down inflation-related phenomena in a technical style, even referencing the channels of monetary policy that are thought to be at work. In such articles, causal chains and forks may intermix, even within a single sentence, yielding multiple narratives from a short text span. When \texttt{GPT-4o} struggles with disentangling these structures and coding all narratives separately, the model's narrative count gets notably deflated for the respective article. Conversely, narrative-sparse articles tend to focus less on textbook economic explanations and more on human interest and storytelling. Given the latter, the model tends to infer inaccurate or speculated narratives much more liberally than human coders. This lack of consistency is related to the phenomenon of hallucinations, a term describing the well-known tendency of LLMs to generate plausible but inaccurate answers \parencite{Huang2023a}.

In general, it appears that the model performs particularly well on short documents with relatively straightforward narratives, while struggling with longer, more complex documents. \texttt{GPT-4o} sometimes fails to detect crucial narrative structures, especially when narratives are implicitly stated or span multiple sentences. As the performance metrics in \autoref{tab:performance} illustrate, these characteristics come with substantial downstream effects: \texttt{GPT-4o} averages 1.25 major deviations per document overall, more than twice the rate of any expert coder. This translate to an \enquote{unexpected major-deviation rate} of 0.91 narratives per document--—a metric that captures how many more major deviation the model made per case compared to the human experts. Taken together, these false positives and false negatives drive down the model’s overall accuracy to 44\%, compared to the 67–74\% accuracy range achieved by individual annotators.

\begin{table}[b]
\centering
\begin{tabular}{lcccc}
\toprule
\textbf{Measure} & \textbf{Expert 1} & \textbf{Expert 2} & \textbf{Expert 3} & \textbf{Model} \\
\midrule
Major-deviation rate             & 0.40 & 0.35 & 0.49 & 1.25 \\
Unexpected major deviations      & \textemdash & \textemdash & \textemdash & 0.91 \\
Accuracy (vs.\ gold)             & 0.72 & 0.74 & 0.67 & 0.44 \\
Jaccard similarity               & 0.59 & 0.60 & 0.59 & 0.40 \\
\bottomrule
\end{tabular}
\vspace{0.5em}
\caption{Annotator agreement and deviation metrics for human experts and \texttt{GPT-4o}.}
\label{tab:performance}
\end{table}

It is important to note that not all errors that we classify as major deviations are created equal. Some of the deviations \texttt{GPT4o} produces represent grave misunderstandings and violations of the codebook while others are quite subtle. Occasionally, the economic expertise encoded in the model appears to collide with the aim of faithfully representing the contents of the documents. For example, \texttt{GPT-4o} encodes \enquote{plans to slow an influx of hard currency that is fueling rapid money-supply growth and pushing up inflation} as a narrative chain, asserting that \textit{money-supply growth} is said to \textit{push up inflation}. While economically plausible, the text clearly asserts the \textit{influx of hard currency} as the cause for inflation. While both variants are related, the model's reading shifts the focal point and chain of causality, resulting in a distorted interpretation of the document.

In an additional evaluation, we benchmark the narrative extraction capabilities using a Jaccard similarity metric. This merely quantitative approach quantifies lexical overlap between predicted and reference token sets on a scale from 0 (no overlap) to 1 (perfect match). Matching the impressions described earlier, the human experts outperform the model, achieving mean scores across all documents of J = 0.59 or 0.6, whereas the model reaches J = 0.40. This gap again points towards some limitations, despite the measure being rather lenient in nature, as it rewards partial lexical matches and ignores semantic coherence, causal directionality, or narrative plausibility.

\subsection{Narrative aggregation}

Given the partially promising yet still limited extraction capabilities of \texttt{GPT-4o}, any downstream analysis of the extracted narratives must be interpreted with caution. At this stage, results should be seen as indicative rather than conclusive. Nonetheless, as outlined earlier, our goal is to develop a comprehensive narrative extraction framework which also requires us to consider how the extracted data can be processed and aggregated for future applications and econometric analysis. Below, we present the results of this aggregation step, even though its implications remain exploratory at this point.

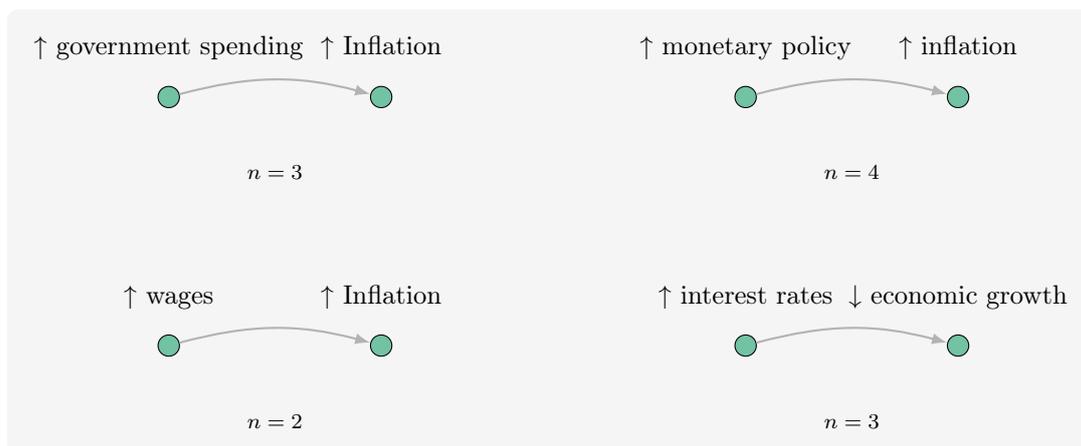
\begin{figure}[h]
  \vspace{1em}
  \centering
  \usetikzlibrary{arrows.meta, positioning}
\definecolor{mint}{HTML}{72C2A4}

\begin{tikzpicture}[
  every node/.style={font=\footnotesize},
  node/.style={circle, draw=black, fill=gray!20, minimum size=8pt, inner sep=0pt},
  highlight/.style={circle, draw=black, fill=mint, minimum size=8pt, inner sep=0pt},
  arrow/.style={-{Latex[length=2mm]}, thick, draw=gray!60}
]

% ROW 1, COL 1
\node (a1) [highlight] at (0,0) {};
\node (a2) [highlight, right=2.5cm of a1] {};
\draw[arrow] (a1) to[bend left=15] (a2);
\node[above=0.2cm of a1] (gs) {↑ government spending};
\node[above=0.2cm of a2] {↑ Inflation};
\node at ($(a1)!0.5!(a2)+(0,-1)$) {\scriptsize $n=3$};

% ROW 1, COL 2
\node (b1) [highlight, right=4.5cm of a2] {};
\node (b2) [highlight, right=2.5cm of b1] {};
\draw[arrow] (b1) to[bend left=15] (b2);
\node[above=0.2cm of b1] {↑ monetary policy };
\node[above=0.2cm of b2] (dv) { ↑ inflation};
\node at ($(b1)!0.5!(b2)+(0,-1)$) {\scriptsize $n=4$};

% ROW 2, COL 1
\node (c1) [highlight, below=3cm of a1] {};
\node (c2) [highlight, right=2.5cm of c1] {};
\draw[arrow] (c1) to[bend left=15] (c2);
\node[above=0.2cm of c1] {↑ wages};
\node[above=0.2cm of c2] {↑ Inflation};
\node at ($(c1)!0.5!(c2)+(0,-1)$) (n21) {\scriptsize $n=2$};

% ROW 2, COL 2
\node (d1) [highlight, right=4.5cm of c2] {};
\node (d2) [highlight, right=2.5cm of d1] {};
\draw[arrow] (d1) to[bend left=15] (d2);
\node[above=0.2cm of d1] {↑ interest rates };
\node[above=0.2cm of d2] (eg) { ↓ economic growth};
\node at ($(d1)!0.5!(d2)+(0,-1)$) (n22) {\scriptsize $n=3$};

\begin{scope}[on background layer]
  \node[fill=gray!8, rounded corners=4pt, inner sep=6pt,
        fit=(gs) (n21) (dv) (eg)] {};
\end{scope}

\end{tikzpicture}
  \caption{Simplified narratives derived from our LLM based narrative extraction pipeline. "↑" in the context of monetary policy translates to \enquote{loose m.p.}. n=x indicates how often this narrative can be found in our sample.}
  \label{fig:nars_visualized}
\end{figure}

The abstraction and harmonization of extracted events enable us to compare and group narrative elements across documents. This, in turn, allows for the identification of frequently recurring causal patterns, such as the link between government spending and inflation, or between interest rate hikes and reduced economic growth. \autoref{fig:nars_visualized} illustrates a selection of such simplified narrative arcs that appeared multiple times across our corpus.

One narrative that recurs notably in our sample is the link between government spending and inflation (n=3). This causal association aligns with theoretical arguments suggesting that expansionary fiscal policy---especially if debt-financed---can trigger inflationary pressure, particularly in environments of constrained monetary policy or supply-side rigidity \parencite{sargent1981some}. Notably, such narratives are not only central to formal macroeconomic models but also resonate in journalistic discourse \parencite{Andre2022, schmidt2025narrating}.

If such narratives appear frequently in the media, they are expected to influence how individuals interpret macroeconomic developments, especially in periods of heightened uncertainty. As discussed in Section~\ref{chap2}, narratives are not just post-hoc explanations, but sense-making stories that shape how people form expectations and make decisions. This might certainly also apply to the kinds of narratives presented here. For instance, if media audiences are repeatedly exposed to a narrative like \enquote{higher wages cause rising inflation} (n=2 in our sample), it is reasonable to assume that such frames may shape their thinking, e.g., influencing how they approach future wage negotiations. Although recent research points in this direction \parencite{Andre2022}, it remains unclear how often a reader needs to encounter a given narrative for it to have a measurable effect. With most prominent narrative types in our sample occurring in just 2–4 of 80 articles, the strength and reach of such effects remain speculative and warrant further investigation.

Other frequently recurring narratives include the effect of loose monetary policy on inflation rates (n=4)---which is also a prominent argument in both theory and public discussion---, and interest rate hikes on economic growth (n=3). All four of these narrative arcs point toward classical macroeconomic concerns that have been widely discussed in both academic and policy debates.
%The fact that these narratives emerge multiple times in our relatively small evaluation corpus suggests that the LLM-based extraction pipeline is capable of identifying economically meaningful causal claims.
The most frequently occurring narrative-parts (valence-topic combinations) in our sample are presented in \autoref{tab:valence_topic}.

\begin{table}[ht]
\vspace{1em}
\centering
\begin{tabular}{lccl c}
\toprule
\textbf{Event (example)} & \textbf{Valence} & \textbf{Sym.} & \textbf{Topic} & \textbf{Count} \\
\midrule
higher inflation & rising  & ↑ & Inflation        & 58 \\
economic stability in the us & positive  & ↑ & Economy   & 29 \\
higher interest rates & rising  & ↑ & Interest rates   & 27 \\
the Fed's aggressiveness & loose   & ↑ & Monetary policy  & 25 \\
the ECB tightening its m.p. & tight   & ↓ & Monetary policy  & 23 \\
economy plunges toward hard landing & falling  & ↓ & Economy   & 22 \\
lower inflation rates & falling & ↓ & Inflation        & 21 \\
higher gas prices & rising  & ↑ & Energy prices    & 19 \\
stocks are getting attractive again & rising & ↑ & Stock Market     &  15 \\
\bottomrule
\end{tabular}
\vspace{0.5em}
\caption{Most frequent valence-topic combinations in extracted narratives. The presented \textit{Events} (left column) are sampled from all events that translate to the respective valence-topic combination.}
\label{tab:valence_topic}
\end{table}

Further analysis of the directional roles played by different events in the extracted narratives reveals additional information on the structure of economic storytelling in the media. For instance, the topic \textit{stock market} appears predominantly as an \enquote{effect event} (90\% of all instances) suggesting that (business) journalism often frames stock prices as a barometer reacting to other economic developments. In contrast, narratives in which stock market movements are framed as a causal force influencing other domains are relatively rare. Conversely, \textit{government spending} is predominantly discussed as a \enquote{cause event} (91\% of instances) highlighting its frequent use as an explanatory device in inflation reporting. This asymmetry mirrors typical theoretical priors in economics: while public spending is often viewed as a policy lever that sets economic processes in motion, the stock market tends to be treated as an outcome variable, shaped by shifts in policy, sentiment, or macroeconomic conditions.

However, the aggregation process also reveals important limitations. Despite careful normalization and abstraction, many extracted narratives remain highly context-specific and cannot be assigned to larger clusters. This is partly due to the large combinatorial space of possible valence-topic pairs: the number of possible combinations of clusters A and B and valences A and B is very high, leading to overall rare instances where different texts lead to the same simplified narrative statement. After decomposition and standardization, our 80 evaluation documents yielded a total of 409 distinct narratives, all made up of two valences and two events. Many of them only appear once in the whole dataset, which explains why we count only two or three instances of our most prominent complete narratives.

This lack of convergence, however, is not considered as a methodological weakness. Instead, it reflects a central feature of narratives as they appear in public discourse, and also in the media. Most newspaper articles tend to convey causal claims that are tailored to specific events, actors, and settings, making generalization inherently difficult. Additionally, our strict definition of narratives as precise, two-event causal claims may contribute to this fragmentation, as it emphasizes nuance and specificity over generalization. It is also important to note that not all causal relationships extracted by the model are economically meaningful. Some narratives form around very specific events, locations or protagonists, which cannot be translated to a broader economic category and may not even be of notable relevance.

All in all, our results highlight both the potential and the limitations of LLM-based narrative aggregation and clustering. While the described framework is capable of detecting economically plausible causal structures and producing interpretable output, further methodological refinement is needed to move from fragmented extractions toward more robust generalizations.

\section{Discussion}
\label{chap7}

Our study provides evidence that general-purpose language models can be meaningfully used to extract economic narratives from text. The results achieved by \texttt{GPT-4o} on our complex annotation task are promising in that they show considerable improvement compared to earlier attempts using previous generations of LLMs. Identifying economic narratives from text requires abstract reasoning, domain-specific knowledge, and the ability to resolve ambiguity and co-reference. We show that a general-purpose, instruction-tuned LLM is able to approach this task with some consistency without any task-specific fine-tuning. This illustrates meaningful development in the field of narrative economics.
%In prior evaluations conducted one year earlier, model outputs were considerably less accurate and less consistent.
The progress we observe here reflects ongoing advances in language model capabilities, particularly in tasks that require instruction-following and context-sensitive reasoning.

We also see considerable promise in combining LLM-based extraction with scalable methods of abstraction and clustering. Our initial experiments show that harmonizing extracted narratives into macroeconomic categories enables the identification of recurring narrative patterns. Although most extracted narratives remain too specific to be aggregated meaningfully, the potential for scalable aggregation is clear. Recent work by Reccius (2025, forthcoming), for example, introduces a polar embedding-based approach that could be used to measure the distance between narratives along theoretically meaningful dimensions. Integrating such techniques with LLM-based extraction appears promising for both theoretical and empirical work on economic narratives.

Our findings also reflect some critical shortcomings of LLMs. While tracing the exact origins of each LLM-misstep is currently not possible, it is worth remembering that LLMs are pre-trained to produce coherent text and then fine-tuned to follow instructions. There is some evidence that, as a side effect of this training, models occasionally try to oversatisfy prompts, placing helpfulness over accuracy and thus avoiding rejections \parencite{chen2025helpfulness}. This tendency results in the model trying to 'squeeze blood from a stone', yielding outputs that are unfaithful to instructions. These drawbacks are especially problematic in research settings and must be addressed.

 %At the same time, our findings underscore some limitations of current LLMs. Most notably, \texttt{GPT-4o} exhibits some biases in narrative density, at least when prompted in the presented way, under-identifying narratives in some contexts while over-identifying them in others. The model struggles with structurally complex narratives, particularly those involving forks or chains. It also shows limited adaptability when encountering highly challenging edge cases or implicit causal links between events that may span multiple sentences. These shortcomings suggest that, while LLMs possess strong general capabilities, economic narrative extraction is a specialized task where expert judgment cannot (yet) be replaced by general-purpose LLMs. 

Importantly, our evaluation approach also highlights the challenges of defining and operationalizing the concept of an economic narrative in practice. Even among expert annotators, we observed substantial variation in annotator opinion about which narratives were identified, how events were segmented, and how causal relations were interpreted. This variability stresses the subjective nature of the sense-making process that occurs when people consume narratives through media. The extraction of economic narratives from text is inherently interpretive, even when formal definitions and structured codebooks are applied. Consequently, fully automating the process remains an ambitious goal.

Looking ahead, we envision a future in which economic narrative research will increasingly be shaped by hybrid workflows—pairing the scalability of language models with the interpretive capacity of human experts. Rather than replacing researchers, LLMs will act as amplifiers of their judgment. Human coders will remain central in defining what constitutes a narrative in the first place, developing precise annotation frameworks, and crafting high-quality prompts that translate conceptual definitions into machine-readable tasks. One key component of this collaboration will be the creation of exemplary few-shot examples, which LLMs rely on for task adaptation.

Accordingly, the time saved by automating the annotation of thousands of documents will not come without cost. It will need to be reinvested in the evaluation process itself. Unlike more standardized tasks such as sentiment classification, narrative extraction requires a deep understanding of semantics, context, and domain-specific framing to evaluate if a models' results are valid. That is, quick eyeballing or resorting to simple accuracy scores will not suffice. Instead, each output must be assessed on its interpretive merit, demanding a level of scrutiny that (for now) only human coders can provide. In that sense, our study underscores the continued importance of human involvement in content analysis and social science research \parencite{haim2023re}. LLMs may extend what is possible; but only when embedded in workflows that preserve interpretive oversight and methodological rigor.

To conclude, our results suggest that while narrative extraction via LLMs is still in its early stages, it offers potential for scalable, interpretable, and theoretically informed analysis of economic discourse. With further refinement, including improved prompts, fine-tuned models, and downstream validation, the approach has the potential to contribute meaningfully to the emerging field of narrative economics.

\subsection{Outlook}
While LLMs will further develop and excel even more at language understanding in the future, we believe that the performance of the current generation of models can be improved on with more specialized training, such as (parameter-efficient) fine-tuning. As best practices for prompting evolve alongside model architectures, we see a need to continuously refine both our extraction and aggregation techniques.

The aggregation framework presented in this paper reflects only an initial stage. Future work should explore more advanced methods of narrative clustering and abstraction. A promising direction is offered by Reccius (2025, forthcoming), who proposes a polar embedding-based approach to improve the alignment of semantically similar narrative elements. Incorporating such techniques may allow us to systematically group related narratives across documents, even when surface-level variation is high. Once greater reliability in extraction and aggregation is achieved, a natural next step is to scale the method to larger corpora. This would enable a more systematic assessment of narrative prevalence and diffusion over time and across domains. 

However, while the performance of \texttt{GPT-4o} on our task showed that Large Language Models have great potential for narrative extraction, using them requires good hardware or financial backing. In addition to that, the environmental impact of running such models should not be understated. We therefore also aim to develop methods that cleverly combine scalable NLP methods with LLMs, enabling a thorough analysis of large corpora with a minimal amount of resources.

Ultimately, this paper focused on establishing the feasibility of LLM-based narrative extraction. Future research should move toward substantive applications. In particular, linking extracted narratives to external economic indicators—such as inflation expectations, consumer sentiment, or financial market volatility—could offer novel insights into the role of narrative framing in macroeconomic dynamics. Such connections would not only extend the methodological contribution of this paper, but also advance our theoretical understanding of how economic stories shape the economy itself.

\section{Conclusion}
\label{chap8}
Narratives play an crucial role in everyday economic decision-making. All actors in an economy, regardless of their level of sophistication or macroeconomic importance, are influenced by the economic narratives circulating in their environment. Extracting such narratives from mass media enables us to trace their origins, understand how they spread through society, and examine their development over time. For economists, this is a crucial yet complex task.

To enable a quantitative analysis of economic narratives, we present an annotation codebook as a stepping stone into extracting economic narratives from a corpus of texts. We utilize this codebook to let three expert annotators code narratives from a total of 100 documents and form a gold-standard data set out of these coded narratives. We then prompt \texttt{GPT-4o} to perform the same task and evaluate it on our gold-standard annotations. 

We categorize deviations from the gold-standard annotations into minor and major deviations, where minor deviations cover formal or small subjective deviations and major deviations cover cases in which a narrative has been coded incorrectly. We also compare the resulting deviations to the ones that resulted from our expert annotators.

Our results indicate that the level of language comprehension and economic expertise encoded in contemporary Large Language Models (LLMs) enables them to extract economic narratives from newspaper articles effectively. We further find that few-shot Chain-of-Thought prompting\textemdash a state-of-the-art AI strategy\textemdash is well-suited for LLM-based text retrieval tasks in economics, as it combines high performance with strong transparency in complex annotation tasks.  

Our findings also underscore some limitations. Most notably, \texttt{GPT-4o} exhibits biases in narrative density, at least when prompted as presented, under-identifying narratives in some contexts while over-identifying them in others. The model struggles with structurally complex narratives, particularly those involving nuanced causal structures like forks or chains. It also shows limited adaptability when encountering challenging edge cases or hidden and implicit causal links between events that may span multiple sentences. 

These shortcomings suggest that, while LLMs possess strong general capabilities, economic narrative extraction is a specialized task where expert judgment cannot (and should not) be fully replaced by general-purpose LLMs. Achieving human-level reliability will require more capable foundational models and methodological refinements. 

\section*{Acknowledgments} %e.g.
This study is part of a project of the Dortmund Center for data-based Media Analysis (DoCMA) at TU Dortmund University and the Narrative Economic Alliance Ruhr (NEAR) project, supported by the Mercator Research Center Ruhr (MERCUR) with project number Ko-2022-0015. It was also partially funded by the Reality Check incubator project at the Research Center for Trustworthy Data Science and Security.

\newpage
\printbibliography[title=References]

@techreport{Flynn2022,
  author      = {Joel P. Flynn and Karthik A. Sastry},
  date        = {2024},
  institution = {National Bureau of Economic Research},
  type   = {NBER Working Paper},
  number = {32602},
  title       = {The Macroeconomics of Narratives},
  doi         = {10.3386/w32602},
  year        = {2024},
}

@article{chen2025helpfulness,
  title={When Helpfulness Backfires: LLMs and the Risk of Misinformation Due to Sycophantic Behavior},
  author={Chen, Shan and Gao, Mingye and Sasse, Kuleen and Hartvigsen, Thomas and Anthony, Brian and Fan, Lizhou and Aerts, Hugo and Gallifant, Jack and Bitterman, Danielle S},
  journal={Research Square},
  pages={rs--3},
  year={2025}
}

@article{10.1145/3731445,
author = {Zhang, Haopeng and Yu, Philip S. and Zhang, Jiawei},
title = {A Systematic Survey of Text Summarization: From Statistical Methods to Large Language Models},
year = {2025},
publisher = {Association for Computing Machinery},
issn = {0360-0300},
doi = {10.1145/3731445},
note = {Just Accepted},
journal = {ACM Comput. Surv.}
}

@techreport{larsen2019business,
  title={Business cycle narratives},
  author={Larsen, Vegard H and Thorsrud, Leif Anders},
  year={2019},
  type={CESifo Working Paper},
  number={7468},
  doi={10.2139/ssrn.3338822}
}

@book{king2020radical,
  title={Radical uncertainty: Decision-making for an unknowable future},
  author={King, Mervyn and Kay, John},
  year={2020},
  publisher={Hachette UK}
}

@misc{fan2024surveyragmeetingllms,
      title={A Survey on RAG Meeting LLMs: Towards Retrieval-Augmented Large Language Models}, 
      author={Wenqi Fan and Yujuan Ding and Liangbo Ning and Shijie Wang and Hengyun Li and Dawei Yin and Tat-Seng Chua and Qing Li},
      year={2024},
      eprint={2405.06211},
      archivePrefix={arXiv},
      primaryClass={cs.CL},
}

@techreport{schmidt2023inflation,
  title={Inflation perception and the formation of inflation expectations},
  author={Schmidt, Torsten and M{\"u}ller, Henrik and Rieger, Jonas and Schmidt, Tobias and Jentsch, Carsten},
  year={2023},
  type={Ruhr Economic Papers},
  number={1025},
  doi={10.4419/96973191}
}

@techreport{werning2022expectations,
  title={Expectations and the Rate of Inflation},
  author={Werning, Iv{\'a}n},
  year={2022},
  type={NBER Working Paper},
  number={30260},
  doi={10.3386/w30260}
}

@article{juster1972inflation,
  title={Inflation and the Consumer},
  author={Juster, F Thomas and Wachtel, Paul and Hymans, Saul and Duesenberry, James},
  doi={10.2307/2534115},
  journal={Brookings Papers on Economic Activity},
  volume={1972},
  number={1},
  pages={71--121},
  year={1972},
  publisher={JSTOR}
}

@article{burch1975stock,
  title={The stock of consumer durables, inflation, and personal saving decisions},
  author={Burch, Susan W and Werneke, Diane},
  journal={The Review of Economics and Statistics},
  pages={141--154},
  year={1975},
  publisher={JSTOR}
}

@article{bachmann2015inflation,
  title={Inflation expectations and readiness to spend: Cross-sectional evidence},
  author={Bachmann, R{\"u}diger and Berg, Tim O and Sims, Eric R},
  journal={American Economic Journal: Economic Policy},
  volume={7},
  number={1},
  pages={1--35},
  year={2015},
  doi={10.1257/pol.20130292},
  publisher={American Economic Association 2014 Broadway, Suite 305, Nashville, TN 37203-2425}
}

@article{ter2022narrative,
  title={Narrative monetary policy surprises and the media},
  author={Ter Ellen, Saskia and Larsen, Vegard H and Thorsrud, Leif Anders},
  journal={Journal of Money, Credit and Banking},
  volume={54},
  number={5},
  pages={1525--1549},
  year={2022},
  publisher={Wiley Online Library}
}

@article{nyman2021news,
  title={News and narratives in financial systems: exploiting big data for systemic risk assessment},
  author={Nyman, Rickard and Kapadia, Sujit and Tuckett, David},
  journal={Journal of Economic Dynamics and Control},
  volume={127},
  doi={10.1016/j.jedc.2021.104119},
  pages={104119},
  year={2021},
  publisher={Elsevier}
}

@TechReport{tuckett2020monetary,
  author={Tuckett, David and Holmes, Douglas and Pearson, Alice and Chaplin, Graeme},
  title={{Monetary policy and the management of uncertainty: a narrative approach}},
  year=2020,
  institution={Bank of England},
  type={Bank of England working papers},
  url={https://ideas.repec.org/p/boe/boeewp/0870.html},
  number={870},
  doi={},
}

@article{conrad2022role,
  title={The role of information and experience for households’ inflation expectations},
  author={Conrad, Christian and Enders, Zeno and Glas, Alexander},
  journal={European Economic Review},
  volume={143},
  pages={104015},
  year={2022},
  doi={10.1016/j.euroecorev.2021.104015},
  publisher={Elsevier}
}

@article{lamla2014role,
  title={The role of media for consumers’ inflation expectation formation},
  author={Lamla, Michael J and Lein, Sarah M},
  journal={Journal of Economic Behavior \& Organization},
  volume={106},
  pages={62--77},
  year={2014},
  publisher={Elsevier}
}

@techreport{schmidt2025narrating,
  title={Narrating inflation: How German economic journalists explain post-covid price rises},
  author={Schmidt, Tobias},
  year={2025},
  type={DoCMA Working Paper},
  number={14},
  doi={10.17877/DE290R-25380}
}

@article{eliaz2020model,
  title={A model of competing narratives},
  author={Eliaz, Kfir and Spiegler, Ran},
  journal={American Economic Review},
  volume={110},
  number={12},
  pages={3786--3816},
  year={2020},
  doi={10.1257/aer.20191099},
  publisher={American Economic Association 2014 Broadway, Suite 305, Nashville, TN 37203}
}

@article{eliaz2024news,
  title={News media as suppliers of narratives (and information)},
  author={Eliaz, Kfir and Spiegler, Ran},
  eprint={2403.09155},
  archivePrefix={arXiv},
  primaryClass={econ.TH},
  year={2024}
}

@article{sims2003implications,
  title={Implications of rational inattention},
  author={Sims, Christopher A},
  journal={Journal of monetary Economics},
  volume={50},
  number={3},
  doi={10.1016/S0304-3932(03)00029-1},
  pages={665--690},
  year={2003},
  publisher={Elsevier}
}

@article{reis2006inattentive,
  title={Inattentive producers},
  author={Reis, Ricardo},
  journal={The Review of Economic Studies},
  volume={73},
  number={3},
  pages={793--821},
  doi={10.1111/j.1467-937X.2006.00396.x},
  year={2006},
  publisher={Wiley-Blackwell}
}

@article{coibion2015phillips,
  title={Is the Phillips curve alive and well after all? Inflation expectations and the missing disinflation},
  author={Coibion, Olivier and Gorodnichenko, Yuriy},
  journal={American Economic Journal: Macroeconomics},
  volume={7},
  number={1},
  pages={197--232},
  year={2015},
  publisher={American Economic Association 2014 Broadway, Suite 305, Nashville, TN 37203-2425},
  doi={10.1257/mac.20130306}
}

@article{bracha2025inflation,
  title={Inflation levels and (in) attention},
  author={Bracha, Anat and Tang, Jenny},
  journal={Review of Economic Studies},
  volume={92},
  number={3},
  doi = {10.1093/restud/rdae063},
  pages={1564--1594},
  year={2025},
  publisher={Oxford University Press UK}
}

@misc{openai2024gpt4ocard,
      title={GPT-4o System Card}, 
      author={OpenAI and : and Aaron Hurst and Adam Lerer and Adam P. Goucher and Adam Perelman and Aditya Ramesh and Aidan Clark and AJ Ostrow and Akila Welihinda and Alan Hayes and Alec Radford and Aleksander Mądry and Alex Baker-Whitcomb and Alex Beutel and Alex Borzunov and Alex Carney and Alex Chow and Alex Kirillov and Alex Nichol and Alex Paino and Alex Renzin and Alex Tachard Passos and Alexander Kirillov and Alexi Christakis and Alexis Conneau and Ali Kamali and Allan Jabri and Allison Moyer and Allison Tam and Amadou Crookes and Amin Tootoochian and Amin Tootoonchian and Ananya Kumar and Andrea Vallone and Andrej Karpathy and Andrew Braunstein and Andrew Cann and Andrew Codispoti and Andrew Galu and Andrew Kondrich and Andrew Tulloch and Andrey Mishchenko and Angela Baek and Angela Jiang and Antoine Pelisse and Antonia Woodford and Anuj Gosalia and Arka Dhar and Ashley Pantuliano and Avi Nayak and Avital Oliver and Barret Zoph and Behrooz Ghorbani and Ben Leimberger and Ben Rossen and Ben Sokolowsky and Ben Wang and Benjamin Zweig and Beth Hoover and Blake Samic and Bob McGrew and Bobby Spero and Bogo Giertler and Bowen Cheng and Brad Lightcap and Brandon Walkin and Brendan Quinn and Brian Guarraci and Brian Hsu and Bright Kellogg and Brydon Eastman and Camillo Lugaresi and Carroll Wainwright and Cary Bassin and Cary Hudson and Casey Chu and Chad Nelson and Chak Li and Chan Jun Shern and Channing Conger and Charlotte Barette and Chelsea Voss and Chen Ding and Cheng Lu and Chong Zhang and Chris Beaumont and Chris Hallacy and Chris Koch and Christian Gibson and Christina Kim and Christine Choi and Christine McLeavey and Christopher Hesse and Claudia Fischer and Clemens Winter and Coley Czarnecki and Colin Jarvis and Colin Wei and Constantin Koumouzelis and Dane Sherburn and Daniel Kappler and Daniel Levin and Daniel Levy and David Carr and David Farhi and David Mely and David Robinson and David Sasaki and Denny Jin and Dev Valladares and Dimitris Tsipras and Doug Li and Duc Phong Nguyen and Duncan Findlay and Edede Oiwoh and Edmund Wong and Ehsan Asdar and Elizabeth Proehl and Elizabeth Yang and Eric Antonow and Eric Kramer and Eric Peterson and Eric Sigler and Eric Wallace and Eugene Brevdo and Evan Mays and Farzad Khorasani and Felipe Petroski Such and Filippo Raso and Francis Zhang and Fred von Lohmann and Freddie Sulit and Gabriel Goh and Gene Oden and Geoff Salmon and Giulio Starace and Greg Brockman and Hadi Salman and Haiming Bao and Haitang Hu and Hannah Wong and Haoyu Wang and Heather Schmidt and Heather Whitney and Heewoo Jun and Hendrik Kirchner and Henrique Ponde de Oliveira Pinto and Hongyu Ren and Huiwen Chang and Hyung Won Chung and Ian Kivlichan and Ian O'Connell and Ian O'Connell and Ian Osband and Ian Silber and Ian Sohl and Ibrahim Okuyucu and Ikai Lan and Ilya Kostrikov and Ilya Sutskever and Ingmar Kanitscheider and Ishaan Gulrajani and Jacob Coxon and Jacob Menick and Jakub Pachocki and James Aung and James Betker and James Crooks and James Lennon and Jamie Kiros and Jan Leike and Jane Park and Jason Kwon and Jason Phang and Jason Teplitz and Jason Wei and Jason Wolfe and Jay Chen and Jeff Harris and Jenia Varavva and Jessica Gan Lee and Jessica Shieh and Ji Lin and Jiahui Yu and Jiayi Weng and Jie Tang and Jieqi Yu and Joanne Jang and Joaquin Quinonero Candela and Joe Beutler and Joe Landers and Joel Parish and Johannes Heidecke and John Schulman and Jonathan Lachman and Jonathan McKay and Jonathan Uesato and Jonathan Ward and Jong Wook Kim and Joost Huizinga and Jordan Sitkin and Jos Kraaijeveld and Josh Gross and Josh Kaplan and Josh Snyder and Joshua Achiam and Joy Jiao and Joyce Lee and Juntang Zhuang and Justyn Harriman and Kai Fricke and Kai Hayashi and Karan Singhal and Katy Shi and Kavin Karthik and Kayla Wood and Kendra Rimbach and Kenny Hsu and Kenny Nguyen and Keren Gu-Lemberg and Kevin Button and Kevin Liu and Kiel Howe and Krithika Muthukumar and Kyle Luther and Lama Ahmad and Larry Kai and Lauren Itow and Lauren Workman and Leher Pathak and Leo Chen and Li Jing and Lia Guy and Liam Fedus and Liang Zhou and Lien Mamitsuka and Lilian Weng and Lindsay McCallum and Lindsey Held and Long Ouyang and Louis Feuvrier and Lu Zhang and Lukas Kondraciuk and Lukasz Kaiser and Luke Hewitt and Luke Metz and Lyric Doshi and Mada Aflak and Maddie Simens and Madelaine Boyd and Madeleine Thompson and Marat Dukhan and Mark Chen and Mark Gray and Mark Hudnall and Marvin Zhang and Marwan Aljubeh and Mateusz Litwin and Matthew Zeng and Max Johnson and Maya Shetty and Mayank Gupta and Meghan Shah and Mehmet Yatbaz and Meng Jia Yang and Mengchao Zhong and Mia Glaese and Mianna Chen and Michael Janner and Michael Lampe and Michael Petrov and Michael Wu and Michele Wang and Michelle Fradin and Michelle Pokrass and Miguel Castro and Miguel Oom Temudo de Castro and Mikhail Pavlov and Miles Brundage and Miles Wang and Minal Khan and Mira Murati and Mo Bavarian and Molly Lin and Murat Yesildal and Nacho Soto and Natalia Gimelshein and Natalie Cone and Natalie Staudacher and Natalie Summers and Natan LaFontaine and Neil Chowdhury and Nick Ryder and Nick Stathas and Nick Turley and Nik Tezak and Niko Felix and Nithanth Kudige and Nitish Keskar and Noah Deutsch and Noel Bundick and Nora Puckett and Ofir Nachum and Ola Okelola and Oleg Boiko and Oleg Murk and Oliver Jaffe and Olivia Watkins and Olivier Godement and Owen Campbell-Moore and Patrick Chao and Paul McMillan and Pavel Belov and Peng Su and Peter Bak and Peter Bakkum and Peter Deng and Peter Dolan and Peter Hoeschele and Peter Welinder and Phil Tillet and Philip Pronin and Philippe Tillet and Prafulla Dhariwal and Qiming Yuan and Rachel Dias and Rachel Lim and Rahul Arora and Rajan Troll and Randall Lin and Rapha Gontijo Lopes and Raul Puri and Reah Miyara and Reimar Leike and Renaud Gaubert and Reza Zamani and Ricky Wang and Rob Donnelly and Rob Honsby and Rocky Smith and Rohan Sahai and Rohit Ramchandani and Romain Huet and Rory Carmichael and Rowan Zellers and Roy Chen and Ruby Chen and Ruslan Nigmatullin and Ryan Cheu and Saachi Jain and Sam Altman and Sam Schoenholz and Sam Toizer and Samuel Miserendino and Sandhini Agarwal and Sara Culver and Scott Ethersmith and Scott Gray and Sean Grove and Sean Metzger and Shamez Hermani and Shantanu Jain and Shengjia Zhao and Sherwin Wu and Shino Jomoto and Shirong Wu and Shuaiqi and Xia and Sonia Phene and Spencer Papay and Srinivas Narayanan and Steve Coffey and Steve Lee and Stewart Hall and Suchir Balaji and Tal Broda and Tal Stramer and Tao Xu and Tarun Gogineni and Taya Christianson and Ted Sanders and Tejal Patwardhan and Thomas Cunninghman and Thomas Degry and Thomas Dimson and Thomas Raoux and Thomas Shadwell and Tianhao Zheng and Todd Underwood and Todor Markov and Toki Sherbakov and Tom Rubin and Tom Stasi and Tomer Kaftan and Tristan Heywood and Troy Peterson and Tyce Walters and Tyna Eloundou and Valerie Qi and Veit Moeller and Vinnie Monaco and Vishal Kuo and Vlad Fomenko and Wayne Chang and Weiyi Zheng and Wenda Zhou and Wesam Manassra and Will Sheu and Wojciech Zaremba and Yash Patil and Yilei Qian and Yongjik Kim and Youlong Cheng and Yu Zhang and Yuchen He and Yuchen Zhang and Yujia Jin and Yunxing Dai and Yury Malkov},
      year={2024},
      eprint={2410.21276},
      archivePrefix={arXiv},
      primaryClass={cs.CL},
}

@inproceedings{brown2020language,
author = {Brown, Tom B. and Mann, Benjamin and Ryder, Nick and Subbiah, Melanie and Kaplan, Jared and Dhariwal, Prafulla and Neelakantan, Arvind and Shyam, Pranav and Sastry, Girish and Askell, Amanda and Agarwal, Sandhini and Herbert-Voss, Ariel and Krueger, Gretchen and Henighan, Tom and Child, Rewon and Ramesh, Aditya and Ziegler, Daniel M. and Wu, Jeffrey and Winter, Clemens and Hesse, Christopher and Chen, Mark and Sigler, Eric and Litwin, Mateusz and Gray, Scott and Chess, Benjamin and Clark, Jack and Berner, Christopher and McCandlish, Sam and Radford, Alec and Sutskever, Ilya and Amodei, Dario},
title = {Language models are few-shot learners},
year = {2020},
isbn = {9781713829546},
publisher = {Curran Associates Inc.},
booktitle = {Proceedings of the 34th International Conference on Neural Information Processing Systems},
articleno = {159},
numpages = {25},
doi={10.5555/3495724.3495883},
series = {NIPS '20}
}

@article{Shiller17,
  title = {Narrative {{Economics}}},
  author = {Shiller, Robert J.},
  year = {2017},
  journal = {American Economic Review},
  volume = {107},
  number = {4},
  pages = {967--1004},
  issn = {0002-8282},
  doi = {10.1257/aer.107.4.967},
  langid = {english}
}

@inproceedings{lange2025narrative,
author="Lange, Kai-Robin and Schmidt, Tobias and Reccius, Matthias and Müller, Henrik and Roos, Michael and Jentsch, Carsten",
title="Narrative Shift detection: A hybrid approach of Dynamic Topic Models and Large Language Models",
booktitle="Proceedings of the Text2Story'25 Workshop",
year = "2025",
url = {https://www.di.ubi.pt/~jpaulo/Text2Story2025/paper6.pdf}
}

@inproceedings{lange22b,
	title = {Zeitenwenden: {Detecting} changes in the {German} political discourse},
	author = {Lange, Kai-Robin and Rieger, Jonas and Benner, Niklas and Jentsch, Carsten},
	year = {2022},
    booktitle = "Proceedings of the 2nd Workshop on Computational Linguistics for the Political and Social Sciences",
    url = {https://old.gscl.org/media/pages/arbeitskreise/cpss/cpss-2022/workshop-proceedings-2022/254133848-1662996909/cpss-2022-proceedings.pdf},
    pages = {47--53}
}

@techreport{lange22a,
	title = {Towards {Extracting} {Collective} {Economic} {Narratives} from {Texts}},
	doi = {10.4419/96973127},
	publisher = {RWI},
    type={Ruhr Economic Papers},
	author = {Lange, Kai-Robin and Reccius, Matthias and Schmidt, Tobias and Müller, Henrik and Roos, Michael and Jentsch, Carsten},
	year = {2022},
    number={963},
}

@misc{GPT4oBenchmark,
  title = {Putting {{GPT-4o}} to the {{Sword}}: {{A Comprehensive Evaluation}} of {{Language}}, {{Vision}}, {{Speech}}, and {{Multimodal Proficiency}}},
  shorttitle = {Putting {{GPT-4o}} to the {{Sword}}},
  author = {Shahriar, Sakib and Lund, Brady and Mannuru, Nishith Reddy and Arshad, Muhammad Arbab and Hayawi, Kadhim and Bevara, Ravi Varma Kumar and Mannuru, Aashrith and Batool, Laiba},
  year = {2024},
  langid = {english},
  eprint={2407.09519v1},
  archivePrefix={arXiv}
}

@article{Bornheim_Grieger_Blaneck_Bialonski_2024, title={Speaker Attribution in German Parliamentary Debates with QLoRA-adapted Large Language Models}, volume={37}, DOI={10.21248/jlcl.37.2024.244}, number={1}, journal={Journal for Language Technology and Computational Linguistics}, author={Bornheim, Tobias and Grieger, Niklas and Blaneck , Patrick Gustav and Bialonski, Stephan}, year={2024}, pages={1–13} }

@article{Gilardi_2023,
   title={ChatGPT outperforms crowd workers for text-annotation tasks},
   volume={120},
   ISSN={1091-6490},
   DOI={10.1073/pnas.2305016120},
   number={30},
   journal={Proceedings of the National Academy of Sciences},
   publisher={Proceedings of the National Academy of Sciences},
   author={Gilardi, Fabrizio and Alizadeh, Meysam and Kubli, Maël},
   year={2023},}

@misc{Llama3,
  title = {The {{Llama}} 3 {{Herd}} of {{Models}}},
  author = {Dubey, Abhimanyu and Jauhri, Abhinav and Pandey, Abhinav and Kadian, Abhishek and {Al-Dahle}, Ahmad and Letman, Aiesha and Mathur, Akhil and Schelten, Alan and Yang, Amy and Fan, Angela and Goyal, Anirudh and Hartshorn, Anthony and Yang, Aobo and Mitra, Archi and Sravankumar, Archie and Korenev, Artem and Hinsvark, Arthur and Rao, Arun and Zhang, Aston and Rodriguez, Aurelien and Gregerson, Austen and Spataru, Ava and Roziere, Baptiste and Biron, Bethany and Tang, Binh and Chern, Bobbie and Caucheteux, Charlotte and Nayak, Chaya and Bi, Chloe and Marra, Chris and McConnell, Chris and Keller, Christian and Touret, Christophe and Wu, Chunyang and Wong, Corinne and Ferrer, Cristian Canton and Nikolaidis, Cyrus and Allonsius, Damien and Song, Daniel and Pintz, Danielle and Livshits, Danny and Esiobu, David and Choudhary, Dhruv and Mahajan, Dhruv and {Garcia-Olano}, Diego and Perino, Diego and Hupkes, Dieuwke and Lakomkin, Egor and AlBadawy, Ehab and Lobanova, Elina and Dinan, Emily and Smith, Eric Michael and Radenovic, Filip and Zhang, Frank and Synnaeve, Gabriel and Lee, Gabrielle and Anderson, Georgia Lewis and Nail, Graeme and Mialon, Gregoire and Pang, Guan and Cucurell, Guillem and Nguyen, Hailey and Korevaar, Hannah and Xu, Hu and Touvron, Hugo and Zarov, Iliyan and Ibarra, Imanol Arrieta and Kloumann, Isabel and Misra, Ishan and Evtimov, Ivan and Copet, Jade and Lee, Jaewon and Geffert, Jan and Vranes, Jana and Park, Jason and Mahadeokar, Jay and Shah, Jeet and {van der Linde}, Jelmer and Billock, Jennifer and Hong, Jenny and Lee, Jenya and Fu, Jeremy and Chi, Jianfeng and Huang, Jianyu and Liu, Jiawen and Wang, Jie and Yu, Jiecao and Bitton, Joanna and Spisak, Joe and Park, Jongsoo and Rocca, Joseph and Johnstun, Joshua and Saxe, Joshua and Jia, Junteng and Alwala, Kalyan Vasuden and Upasani, Kartikeya and Plawiak, Kate and Li, Ke and Heafield, Kenneth and Stone, Kevin and {El-Arini}, Khalid and Iyer, Krithika and Malik, Kshitiz and Chiu, Kuenley and Bhalla, Kunal and {Rantala-Yeary}, Lauren and {van der Maaten}, Laurens and Chen, Lawrence and Tan, Liang and Jenkins, Liz and Martin, Louis and Madaan, Lovish and Malo, Lubo and Blecher, Lukas and Landzaat, Lukas and {de Oliveira}, Luke and Muzzi, Madeline and Pasupuleti, Mahesh and Singh, Mannat and Paluri, Manohar and Kardas, Marcin and Oldham, Mathew and Rita, Mathieu and Pavlova, Maya and Kambadur, Melanie and Lewis, Mike and Si, Min and Singh, Mitesh Kumar and Hassan, Mona and Goyal, Naman and Torabi, Narjes and Bashlykov, Nikolay and Bogoychev, Nikolay and Chatterji, Niladri and Duchenne, Olivier and {\c C}elebi, Onur and Alrassy, Patrick and Zhang, Pengchuan and Li, Pengwei and Vasic, Petar and Weng, Peter and Bhargava, Prajjwal and Dubal, Pratik and Krishnan, Praveen and Koura, Punit Singh and Xu, Puxin and He, Qing and Dong, Qingxiao and Srinivasan, Ragavan and Ganapathy, Raj and Calderer, Ramon and Cabral, Ricardo Silveira and Stojnic, Robert and Raileanu, Roberta and Girdhar, Rohit and Patel, Rohit and Sauvestre, Romain and Polidoro, Ronnie and Sumbaly, Roshan and Taylor, Ross and Silva, Ruan and Hou, Rui and Wang, Rui and Hosseini, Saghar and Chennabasappa, Sahana and Singh, Sanjay and Bell, Sean and Kim, Seohyun Sonia and Edunov, Sergey and Nie, Shaoliang and Narang, Sharan and Raparthy, Sharath and Shen, Sheng and Wan, Shengye and Bhosale, Shruti and Zhang, Shun and Vandenhende, Simon and Batra, Soumya and Whitman, Spencer and Sootla, Sten and Collot, Stephane and Gururangan, Suchin and Borodinsky, Sydney and Herman, Tamar and Fowler, Tara and Sheasha, Tarek and Georgiou, Thomas and Scialom, Thomas and Speckbacher, Tobias and Mihaylov, Todor and Xiao, Tong and Karn, Ujjwal and Goswami, Vedanuj and Gupta, Vibhor and Ramanathan, Vignesh and Kerkez, Viktor and Gonguet, Vincent and Do, Virginie and Vogeti, Vish and Petrovic, Vladan and Chu, Weiwei and Xiong, Wenhan and Fu, Wenyin and Meers, Whitney and Martinet, Xavier and Wang, Xiaodong and Tan, Xiaoqing Ellen and Xie, Xinfeng and Jia, Xuchao and Wang, Xuewei and Goldschlag, Yaelle and Gaur, Yashesh and Babaei, Yasmine and Wen, Yi and Song, Yiwen and Zhang, Yuchen and Li, Yue and Mao, Yuning and Coudert, Zacharie Delpierre and Yan, Zheng and Chen, Zhengxing and Papakipos, Zoe and Singh, Aaditya and Grattafiori, Aaron and Jain, Abha and Kelsey, Adam and Shajnfeld, Adam and Gangidi, Adithya and Victoria, Adolfo and Goldstand, Ahuva and Menon, Ajay and Sharma, Ajay and Boesenberg, Alex and Vaughan, Alex and Baevski, Alexei and Feinstein, Allie and Kallet, Amanda and Sangani, Amit and Yunus, Anam and Lupu, Andrei and Alvarado, Andres and Caples, Andrew and Gu, Andrew and Ho, Andrew and Poulton, Andrew and Ryan, Andrew and Ramchandani, Ankit and Franco, Annie and Saraf, Aparajita and Chowdhury, Arkabandhu and Gabriel, Ashley and Bharambe, Ashwin and Eisenman, Assaf and Yazdan, Azadeh and James, Beau and Maurer, Ben and Leonhardi, Benjamin and Huang, Bernie and Loyd, Beth and De Paola, Beto and Paranjape, Bhargavi and Liu, Bing and Wu, Bo and Ni, Boyu and Hancock, Braden and Wasti, Bram and Spence, Brandon and Stojkovic, Brani and Gamido, Brian and Montalvo, Britt and Parker, Carl and Burton, Carly and Mejia, Catalina and Wang, Changhan and Kim, Changkyu and Zhou, Chao and Hu, Chester and Chu, Ching-Hsiang and Cai, Chris and Tindal, Chris and Feichtenhofer, Christoph and Civin, Damon and Beaty, Dana and Kreymer, Daniel and Li, Daniel and Wyatt, Danny and Adkins, David and Xu, David and Testuggine, Davide and David, Delia and Parikh, Devi and Liskovich, Diana and Foss, Didem and Wang, Dingkang and Le, Duc and Holland, Dustin and Dowling, Edward and Jamil, Eissa and Montgomery, Elaine and Presani, Eleonora and Hahn, Emily and Wood, Emily and Brinkman, Erik and Arcaute, Esteban and Dunbar, Evan and Smothers, Evan and Sun, Fei and Kreuk, Felix and Tian, Feng and Ozgenel, Firat and Caggioni, Francesco and Guzm{\'a}n, Francisco and Kanayet, Frank and Seide, Frank and Florez, Gabriela Medina and Schwarz, Gabriella and Badeer, Gada and Swee, Georgia and Halpern, Gil and Thattai, Govind and Herman, Grant and Sizov, Grigory and Guangyi and Zhang and Lakshminarayanan, Guna and Shojanazeri, Hamid and Zou, Han and Wang, Hannah and Zha, Hanwen and Habeeb, Haroun and Rudolph, Harrison and Suk, Helen and Aspegren, Henry and Goldman, Hunter and Damlaj, Ibrahim and Molybog, Igor and Tufanov, Igor and Veliche, Irina-Elena and Gat, Itai and Weissman, Jake and Geboski, James and Kohli, James and Asher, Japhet and Gaya, Jean-Baptiste and Marcus, Jeff and Tang, Jeff and Chan, Jennifer and Zhen, Jenny and Reizenstein, Jeremy and Teboul, Jeremy and Zhong, Jessica and Jin, Jian and Yang, Jingyi and Cummings, Joe and Carvill, Jon and Shepard, Jon and McPhie, Jonathan and Torres, Jonathan and Ginsburg, Josh and Wang, Junjie and Wu, Kai and U, Kam Hou and Saxena, Karan and Prasad, Karthik and Khandelwal, Kartikay and Zand, Katayoun and Matosich, Kathy and Veeraraghavan, Kaushik and Michelena, Kelly and Li, Keqian and Huang, Kun and Chawla, Kunal and Lakhotia, Kushal and Huang, Kyle and Chen, Lailin and Garg, Lakshya and A, Lavender and Silva, Leandro and Bell, Lee and Zhang, Lei and Guo, Liangpeng and Yu, Licheng and Moshkovich, Liron and Wehrstedt, Luca and Khabsa, Madian and Avalani, Manav and Bhatt, Manish and Tsimpoukelli, Maria and Mankus, Martynas and Hasson, Matan and Lennie, Matthew and Reso, Matthias and Groshev, Maxim and Naumov, Maxim and Lathi, Maya and Keneally, Meghan and Seltzer, Michael L. and Valko, Michal and Restrepo, Michelle and Patel, Mihir and Vyatskov, Mik and Samvelyan, Mikayel and Clark, Mike and Macey, Mike and Wang, Mike and Hermoso, Miquel Jubert and Metanat, Mo and Rastegari, Mohammad and Bansal, Munish and Santhanam, Nandhini and Parks, Natascha and White, Natasha and Bawa, Navyata and Singhal, Nayan and Egebo, Nick and Usunier, Nicolas and Laptev, Nikolay Pavlovich and Dong, Ning and Zhang, Ning and Cheng, Norman and Chernoguz, Oleg and Hart, Olivia and Salpekar, Omkar and Kalinli, Ozlem and Kent, Parkin and Parekh, Parth and Saab, Paul and Balaji, Pavan and Rittner, Pedro and Bontrager, Philip and Roux, Pierre and Dollar, Piotr and Zvyagina, Polina and Ratanchandani, Prashant and Yuvraj, Pritish and Liang, Qian and Alao, Rachad and Rodriguez, Rachel and Ayub, Rafi and Murthy, Raghotham and Nayani, Raghu and Mitra, Rahul and Li, Raymond and Hogan, Rebekkah and Battey, Robin and Wang, Rocky and Maheswari, Rohan and Howes, Russ and Rinott, Ruty and Bondu, Sai Jayesh and Datta, Samyak and Chugh, Sara and Hunt, Sara and Dhillon, Sargun and Sidorov, Sasha and Pan, Satadru and Verma, Saurabh and Yamamoto, Seiji and Ramaswamy, Sharadh and Lindsay, Shaun and Feng, Sheng and Lin, Shenghao and Zha, Shengxin Cindy and Shankar, Shiva and Zhang, Shuqiang and Wang, Sinong and Agarwal, Sneha and Sajuyigbe, Soji and Chintala, Soumith and Max, Stephanie and Chen, Stephen and Kehoe, Steve and Satterfield, Steve and Govindaprasad, Sudarshan and Gupta, Sumit and Cho, Sungmin and Virk, Sunny and Subramanian, Suraj and Choudhury, Sy and Goldman, Sydney and Remez, Tal and Glaser, Tamar and Best, Tamara and Kohler, Thilo and Robinson, Thomas and Li, Tianhe and Zhang, Tianjun and Matthews, Tim and Chou, Timothy and Shaked, Tzook and Vontimitta, Varun and Ajayi, Victoria and Montanez, Victoria and Mohan, Vijai and Kumar, Vinay Satish and Mangla, Vishal and Albiero, V{\'i}tor and Ionescu, Vlad and Poenaru, Vlad and Mihailescu, Vlad Tiberiu and Ivanov, Vladimir and Li, Wei and Wang, Wenchen and Jiang, Wenwen and Bouaziz, Wes and Constable, Will and Tang, Xiaocheng and Wang, Xiaofang and Wu, Xiaojian and Wang, Xiaolan and Xia, Xide and Wu, Xilun and Gao, Xinbo and Chen, Yanjun and Hu, Ye and Jia, Ye and Qi, Ye and Li, Yenda and Zhang, Yilin and Zhang, Ying and Adi, Yossi and Nam, Youngjin and Yu and Wang and Hao, Yuchen and Qian, Yundi and He, Yuzi and Rait, Zach and DeVito, Zachary and Rosnbrick, Zef and Wen, Zhaoduo and Yang, Zhenyu and Zhao, Zhiwei},
  year = {2024},
  langid = {english},
  eprint={2407.21783v2},
  archivePrefix={arXiv}
}

@inproceedings{gueta2024can,
    title = "Can {LLM}s Learn Macroeconomic Narratives from Social Media?",
    author = "Gueta, Almog  and
      Feder, Amir  and
      Gekhman, Zorik  and
      Goldstein, Ariel  and
      Reichart, Roi",
    booktitle = "Findings of the Association for Computational Linguistics: NAACL 2025",
    year = "2025",
    publisher = "Association for Computational Linguistics",
    url = "https://aclanthology.org/2025.findings-naacl.4/",
    pages = "57--78",
    ISBN = "979-8-89176-195-7",
}

@article{ziems2024can,
    title = "Can Large Language Models Transform Computational Social Science?",
    author = "Ziems, Caleb  and
      Held, William  and
      Shaikh, Omar  and
      Chen, Jiaao  and
      Zhang, Zhehao  and
      Yang, Diyi",
    journal = "Computational Linguistics",
    volume = "50",
    number = "1",
    year = "2024",
    publisher = "MIT Press",
    doi = "10.1162/coli_a_00502",
    pages = "237--291"
}

@article{neidhardt2010selbststeuerung,
  title={Selbststeuerung der Wissenschaft: Peer Review},
  author={Neidhardt, Friedhelm},
  journal={Handbuch Wissenschaftspolitik},
  pages={280--292},
  year={2010},
  doi={10.1007/978-3-531-91993-5_19},
  publisher={Springer}
}

@article{burla2008text,
  title={From text to codings: intercoder reliability assessment in qualitative content analysis},
  author={Burla, Laila and Knierim, Birte and Barth, Jurgen and Liewald, Katharina and Duetz, Margreet and Abel, Thomas},
  journal={Nursing research},
  volume={57},
  number={2},
  pages={113--117},
  year={2008},
  publisher={LWW},
  doi = {10.1097/01.NNR.0000313482.33917.7d}
}

@article{mellon2024ais,
author = {Jonathan Mellon and Jack Bailey and Ralph Scott and James Breckwoldt and Marta Miori and Phillip Schmedeman},
title ={Do AIs know what the most important issue is? Using language models to code open-text social survey responses at scale},

journal = {Research \& Politics},
volume = {11},
number = {1},
year = {2024},
doi = {10.1177/20531680241231468},
}

@article{wang2024advanced,
  title={Do advanced language models eliminate the need for prompt engineering in software engineering?},
  author={Wang, Guoqing and Sun, Zeyu and Gong, Zhihao and Ye, Sixiang and Chen, Yizhou and Zhao, Yifan and Liang, Qingyuan and Hao, Dan},
  eprint       = {2411.02093},
  archivePrefix= {arXiv},
  primaryClass = {cs.SE},
  year={2024}
}

@misc{deepseekai2025deepseekr1,
  author       = {DeepSeek-AI},
  title        = {DeepSeek-R1: Incentivizing Reasoning Capability in LLMs via Reinforcement Learning},
  year         = {2025},
  eprint       = {2501.12948},
  archivePrefix= {arXiv},
  primaryClass = {cs.CL},
}

@article{crow2018narratives,
  title={Narratives as tools for influencing policy change},
  author={Crow, Deserai and Jones, Michael},
  journal={Policy \& Politics},
  volume={46},
  number={2},
  pages={217--234},
  doi={10.1332/030557318X15230061022899},
  year={2018},
  publisher={Policy Press}
}

@article{jones2010narrative,
author = {Jones, Michael D. and McBeth, Mark K.},
title = {A Narrative Policy Framework: Clear Enough to Be Wrong?},
journal = {Policy Studies Journal},
volume = {38},
number = {2},
pages = {329-353},
doi = {10.1111/j.1541-0072.2010.00364.x},
year = {2010}
}

@incollection{shanahan2018narrative,
  title={The narrative policy framework},
  author={Shanahan, Elizabeth A and Jones, Michael D and McBeth, Mark K and Radaelli, Claudio M},
  booktitle={Theories of the policy process},
  pages={173--213},
  year={2018},
  publisher={Routledge},
  doi = {10.4324/9780429494284},
  issn={9780429494284}
}

@article{schlaufer2022narrative,
  title={The narrative policy framework: a traveler’s guide to policy stories},
  author={Schlaufer, Caroline and Kuenzler, Johanna and Jones, Michael D and Shanahan, Elizabeth A},
  journal={Politische Vierteljahresschrift},
  volume={63},
  number={2},
  pages={249--273},
  year={2022},
  publisher={Springer},
  doi = {10.1007/s11615-022-00379-6}
}

@inproceedings{Bert,
    title = "{BERT}: Pre-training of Deep Bidirectional Transformers for Language Understanding",
    author = "Devlin, Jacob  and
      Chang, Ming-Wei  and
      Lee, Kenton  and
      Toutanova, Kristina",
    booktitle = "Proceedings of the 2019 Conference of the North {A}merican Chapter of the Association for Computational Linguistics: Human Language Technologies, Volume 1 (Long and Short Papers)",
    year = "2019",
    publisher = "Association for Computational Linguistics",
    doi = "10.18653/v1/N19-1423",
    pages = "4171--4186",
}

@inproceedings{rieger22rolling,
    title = "{RollingLDA}: {A}n Update Algorithm of {L}atent {D}irichlet {A}llocation to Construct Consistent Time Series from Textual Data",
    author = {Rieger, Jonas  and
      Jentsch, Carsten  and
      Rahnenf{\"u}hrer, J{\"o}rg},
    booktitle = "Findings of the Association for Computational Linguistics: EMNLP 2021",
    year = "2021",
    publisher = "Association for Computational Linguistics",
    doi = "10.18653/v1/2021.findings-emnlp.201",
    pages = "2337--2347",
}

@inproceedings{rieger22dynamic,
author={Rieger, Jonas and Lange, Kai-Robin and Flossdorf, Jonathan and Jentsch, Carsten},
year = {2022},
title = {Dynamic change detection in topics based on rolling LDAs},
booktitle = {Proceedings of the Text2Story'22 Workshop},
url = {http://ceur-ws.org/Vol-3117/paper1.pdf},
pages = {5-13}
}

@article{ash21relatio, title={Relatio: Text Semantics Capture Political and Economic Narratives}, volume={32}, DOI={10.1017/pan.2023.8}, number={1}, journal={Political Analysis}, author={Ash, Elliott and Gauthier, Germain and Widmer, Philine}, year={2024}, pages={115–132}}

@techreport{benner22,
author = {Benner, Niklas and Lange, Kai-Robin and Jentsch, Carsten},
doi={10.4419/96973126},
title = {Named Entity Narratives},
year = 2022,
number = 962,
type = {Ruhr Economic Papers},
volume = {962}
}

@misc{o1,
    key = {Learning to reason with LLMs},
    author = {OpenAI},
    url = {https://openai.com/index/learning-to-reason-with-llms/},
    urldate = {2025-05-07},
    year={2024}
}

@article{haim2023re,
  title={(Re) Establishing quality criteria for content analysis: A critical perspective on the field’s core method},
  author={Haim, Mario and Hase, Valerie and Schindler, Johanna and Bachl, Marko and Domahidi, Emese},
  journal={Studies in Communication and Media (SCM)},
  volume={12},
  pages={277--288},
  year={2023}
}

@misc{openai2024gpt4technicalreport,
      title={GPT-4 Technical Report}, 
      author={OpenAI and Josh Achiam and Steven Adler and Sandhini Agarwal and Lama Ahmad and Ilge Akkaya and Florencia Leoni Aleman and Diogo Almeida and Janko Altenschmidt and Sam Altman and Shyamal Anadkat and Red Avila and Igor Babuschkin and Suchir Balaji and Valerie Balcom and Paul Baltescu and Haiming Bao and Mohammad Bavarian and Jeff Belgum and Irwan Bello and Jake Berdine and Gabriel Bernadett-Shapiro and Christopher Berner and Lenny Bogdonoff and Oleg Boiko and Madelaine Boyd and Anna-Luisa Brakman and Greg Brockman and Tim Brooks and Miles Brundage and Kevin Button and Trevor Cai and Rosie Campbell and Andrew Cann and Brittany Carey and Chelsea Carlson and Rory Carmichael and Brooke Chan and Che Chang and Fotis Chantzis and Derek Chen and Sully Chen and Ruby Chen and Jason Chen and Mark Chen and Ben Chess and Chester Cho and Casey Chu and Hyung Won Chung and Dave Cummings and Jeremiah Currier and Yunxing Dai and Cory Decareaux and Thomas Degry and Noah Deutsch and Damien Deville and Arka Dhar and David Dohan and Steve Dowling and Sheila Dunning and Adrien Ecoffet and Atty Eleti and Tyna Eloundou and David Farhi and Liam Fedus and Niko Felix and Simón Posada Fishman and Juston Forte and Isabella Fulford and Leo Gao and Elie Georges and Christian Gibson and Vik Goel and Tarun Gogineni and Gabriel Goh and Rapha Gontijo-Lopes and Jonathan Gordon and Morgan Grafstein and Scott Gray and Ryan Greene and Joshua Gross and Shixiang Shane Gu and Yufei Guo and Chris Hallacy and Jesse Han and Jeff Harris and Yuchen He and Mike Heaton and Johannes Heidecke and Chris Hesse and Alan Hickey and Wade Hickey and Peter Hoeschele and Brandon Houghton and Kenny Hsu and Shengli Hu and Xin Hu and Joost Huizinga and Shantanu Jain and Shawn Jain and Joanne Jang and Angela Jiang and Roger Jiang and Haozhun Jin and Denny Jin and Shino Jomoto and Billie Jonn and Heewoo Jun and Tomer Kaftan and Łukasz Kaiser and Ali Kamali and Ingmar Kanitscheider and Nitish Shirish Keskar and Tabarak Khan and Logan Kilpatrick and Jong Wook Kim and Christina Kim and Yongjik Kim and Jan Hendrik Kirchner and Jamie Kiros and Matt Knight and Daniel Kokotajlo and Łukasz Kondraciuk and Andrew Kondrich and Aris Konstantinidis and Kyle Kosic and Gretchen Krueger and Vishal Kuo and Michael Lampe and Ikai Lan and Teddy Lee and Jan Leike and Jade Leung and Daniel Levy and Chak Ming Li and Rachel Lim and Molly Lin and Stephanie Lin and Mateusz Litwin and Theresa Lopez and Ryan Lowe and Patricia Lue and Anna Makanju and Kim Malfacini and Sam Manning and Todor Markov and Yaniv Markovski and Bianca Martin and Katie Mayer and Andrew Mayne and Bob McGrew and Scott Mayer McKinney and Christine McLeavey and Paul McMillan and Jake McNeil and David Medina and Aalok Mehta and Jacob Menick and Luke Metz and Andrey Mishchenko and Pamela Mishkin and Vinnie Monaco and Evan Morikawa and Daniel Mossing and Tong Mu and Mira Murati and Oleg Murk and David Mély and Ashvin Nair and Reiichiro Nakano and Rajeev Nayak and Arvind Neelakantan and Richard Ngo and Hyeonwoo Noh and Long Ouyang and Cullen O'Keefe and Jakub Pachocki and Alex Paino and Joe Palermo and Ashley Pantuliano and Giambattista Parascandolo and Joel Parish and Emy Parparita and Alex Passos and Mikhail Pavlov and Andrew Peng and Adam Perelman and Filipe de Avila Belbute Peres and Michael Petrov and Henrique Ponde de Oliveira Pinto and Michael and Pokorny and Michelle Pokrass and Vitchyr H. Pong and Tolly Powell and Alethea Power and Boris Power and Elizabeth Proehl and Raul Puri and Alec Radford and Jack Rae and Aditya Ramesh and Cameron Raymond and Francis Real and Kendra Rimbach and Carl Ross and Bob Rotsted and Henri Roussez and Nick Ryder and Mario Saltarelli and Ted Sanders and Shibani Santurkar and Girish Sastry and Heather Schmidt and David Schnurr and John Schulman and Daniel Selsam and Kyla Sheppard and Toki Sherbakov and Jessica Shieh and Sarah Shoker and Pranav Shyam and Szymon Sidor and Eric Sigler and Maddie Simens and Jordan Sitkin and Katarina Slama and Ian Sohl and Benjamin Sokolowsky and Yang Song and Natalie Staudacher and Felipe Petroski Such and Natalie Summers and Ilya Sutskever and Jie Tang and Nikolas Tezak and Madeleine B. Thompson and Phil Tillet and Amin Tootoonchian and Elizabeth Tseng and Preston Tuggle and Nick Turley and Jerry Tworek and Juan Felipe Cerón Uribe and Andrea Vallone and Arun Vijayvergiya and Chelsea Voss and Carroll Wainwright and Justin Jay Wang and Alvin Wang and Ben Wang and Jonathan Ward and Jason Wei and CJ Weinmann and Akila Welihinda and Peter Welinder and Jiayi Weng and Lilian Weng and Matt Wiethoff and Dave Willner and Clemens Winter and Samuel Wolrich and Hannah Wong and Lauren Workman and Sherwin Wu and Jeff Wu and Michael Wu and Kai Xiao and Tao Xu and Sarah Yoo and Kevin Yu and Qiming Yuan and Wojciech Zaremba and Rowan Zellers and Chong Zhang and Marvin Zhang and Shengjia Zhao and Tianhao Zheng and Juntang Zhuang and William Zhuk and Barret Zoph},
      year={2024},
      eprint={2303.08774},
      archivePrefix={arXiv},
      primaryClass={cs.CL},
}

@article{blinder2008central,
  title={Central bank communication and monetary policy: A survey of theory and evidence},
  author={Blinder, Alan S and Ehrmann, Michael and Fratzscher, Marcel and De Haan, Jakob and Jansen, David-Jan},
  journal={Journal of economic literature},
  volume={46},
  doi = {10.1257/jel.46.4.910},
  number={4},
  pages={910--45},
  year={2008}
}

@article{sargent1981some,
  title={Some unpleasant monetarist arithmetic},
  author={Sargent, Thomas J and Wallace, Neil and others},
  journal={Federal reserve bank of minneapolis quarterly review},
  volume={5},
  number={3},
  pages={1--17},
  year={1981}
}

@article{kalamara2020making,
  title={Making text count: economic forecasting using newspaper text},
  author={Kalamara, Eleni and Turrell, Arthur and Redl, Chris and Kapetanios, George and Kapadia, Sujit},
  journal={Journal of Applied Econometrics},
  year={2020},
  publisher={Wiley Online Library}
}

@article{roos2024,
  title = {Narratives in Economics},
  author = {Roos, Michael and Reccius, Matthias},
  year = {2024},
  journal = {Journal of Economic Surveys},
  volume = {38},
  number = {2},
  pages = {303--341},
  doi = {10.1111/joes.12576},
}

@techreport{benabou2018narratives,
  author    = {B{\'e}nabou, Roland and Falk, Armin and Tirole, Jean},
  number    = {24798},
  title     = {Narratives, Imperatives, and Moral Reasoning},
  type      = {NBER Working Paper},
  year      = {2018},
  doi       = {10.3386/w24798},
  date      = {2018},
  publisher = {{National Bureau of Economic Research}},
  series    = {IASS Study},
}

@TechReport{Andre2022,
  author    = {Peter Andre and Ingar Haaland and Christopher Roth and Johannes Wohlfart},
  date      = {2024},
  title     = {Narratives about the Macroeconomy},
  doi       = {10.2139/ssrn.4947636},
  number    = {426},
  type      = {SAFE Working Paper},
  publisher = {Elsevier {BV}},
  year      = {2022},
}

@InProceedings{Mondorf2024,
  author    = {Mondorf, Philipp and Plank, Barbara},
  booktitle = {Proceedings of the 62nd Annual Meeting of the Association for Computational Linguistics (Volume 1: Long Papers)},
  date      = {2024},
  title     = {Comparing Inferential Strategies of Humans and Large Language Models in Deductive Reasoning},
  doi       = {10.18653/v1/2024.acl-long.508},
  pages     = {9370--9402},
  publisher = {Association for Computational Linguistics},
}

@Article{Gurkaynak2005,
  author    = {Refet S. Gurkaynak and Brian P. Sack and Eric T. Swanson},
  title     = {Do Actions Speak Louder Than Words? The Response of Asset Prices to Monetary Policy Actions and Statements},
  doi       = {10.2139/ssrn.633281},
  pages     = {55--93},
  volume    = {1},
  journal   = {International Journal of Central Banking},
  publisher = {Elsevier {BV}},
  year      = {2005},
}

@Article{Hansen2019,
  author    = {Stephen Hansen and Michael McMahon and Matthew Tong},
  title     = {The long-run information effect of central bank communication},
  doi       = {10.1016/j.jmoneco.2019.09.002},
  pages     = {185--202},
  volume    = {108},
  journal   = {Journal of Monetary Economics},
  publisher = {Elsevier {BV}},
  year      = {2019},
}

@Article{Hansen2017,
  author    = {Stephen Hansen and Michael McMahon and Andrea Prat},
  title     = {Transparency and Deliberation Within the {FOMC}: A Computational Linguistics Approach},
  doi       = {10.1093/qje/qjx045},
  number    = {2},
  pages     = {801--870},
  volume    = {133},
  journal   = {The Quarterly Journal of Economics},
  publisher = {Oxford University Press ({OUP})},
  year      = {2017},
}

@Book{Pearl2009,
  author    = {Pearl, Judea},
  date      = {2009},
  title     = {Causality},
  isbn      = {9781139632997},
  publisher = {Cambridge University Press},
  ean       = {9781139632997},
  year      = {2009},
}

@inproceedings{Kojima2022,
author = {Kojima, Takeshi and Gu, Shixiang Shane and Reid, Machel and Matsuo, Yutaka and Iwasawa, Yusuke},
title = {Large language models are zero-shot reasoners},
year = {2022},
doi = {10.5555/3600270.3601883},
isbn = {9781713871088},
publisher = {Curran Associates Inc.},
booktitle = {Proceedings of the 36th International Conference on Neural Information Processing Systems},
articleno = {1613},
numpages = {15},
series = {NIPS '22}
}

@TechReport{Wei2021,
  author      = {Jason Wei and Maarten Bosma and Vincent Y. Zhao and Kelvin Guu and Adams Wei Yu and Brian Lester and Nan Du and Andrew M. Dai and Quoc V. Le},
  date        = {2021-09-03},
  title       = {Finetuned Language Models Are Zero-Shot Learners},
  eprint      = {2109.01652},
  eprintclass = {cs.CL},
  eprinttype  = {arXiv},
}

@Article{Gorodnichenko2023,
  author       = {Yuriy Gorodnichenko and Tho Pham and Oleksandr Talavera},
  date         = {2023-02},
  journaltitle = {American Economic Review},
  title        = {The Voice of Monetary Policy},
  doi          = {10.1257/aer.20220129},
  number       = {2},
  pages        = {548--584},
  volume       = {113},
  publisher    = {American Economic Association},
}

@Article{Liu2023a,
    title = {Lost in the Middle: How Language Models Use Long Contexts},
    author = {Liu, Nelson F.  and
      Lin, Kevin  and
      Hewitt, John  and
      Paranjape, Ashwin  and
      Bevilacqua, Michele  and
      Petroni, Fabio  and
      Liang, Percy},
    journal = {Transactions of the Association for Computational Linguistics},
    volume = {12},
    year = {2024},
    publisher = {MIT Press},
    url = {https://aclanthology.org/2024.tacl-1.9/},
    doi = {10.1162/tacl_a_00638},
    pages = {157--173},
}

@InProceedings{Zhong2025,
  author    = {Zhong, Meizhi and Zhang, Chen and Lei, Yikun and Liu, Xikai and Gao, Yan and Hu, Yao and Chen, Kehai and Zhang, Min},
  booktitle = {Proceedings of the 31st International Conference on Computational Linguistics},
  date      = {2025},
  title     = {Understanding the {R}o{PE} Extensions of Long-Context {LLM}s: An Attention Perspective},
  pages     = {8955--8962},
  publisher = {Association for Computational Linguistics},
  url       = {https://aclanthology.org/2025.coling-main.600},
}

@Article{Song2023,
  author       = {Song, Yisheng and Wang, Ting and Cai, Puyu and Mondal, Subrota K. and Sahoo, Jyoti Prakash},
  date         = {2023-07},
  journaltitle = {ACM Computing Surveys},
  title        = {A Comprehensive Survey of Few-shot Learning: Evolution, Applications, Challenges, and Opportunities},
  doi          = {10.1145/3582688},
  issn         = {1557-7341},
  number       = {13s},
  pages        = {1--40},
  volume       = {55},
  publisher    = {Association for Computing Machinery (ACM)},
}

@misc{Tian2024,
      title={Distance between Relevant Information Pieces Causes Bias in Long-Context LLMs}, 
      author={Runchu Tian and Yanghao Li and Yuepeng Fu and Siyang Deng and Qinyu Luo and Cheng Qian and Shuo Wang and Xin Cong and Zhong Zhang and Yesai Wu and Yankai Lin and Huadong Wang and Xiaojiang Liu},
      year={2024},
      eprint={2410.14641},
      archivePrefix={arXiv},
      primaryClass={cs.CL},
}

@Book{Saldana2016,
  author    = {Saldaña, Johnny},
  title     = {The coding manual for qualitative researchers.},
  edition   = {3E.},
  isbn      = {9781473902480},
  publisher = {SAGE},
  booktitle = {The coding manual for qualitative researchers},
  lccn      = {2015938391},
  year      = {2016},
}

@WWW{DeFiore2025,
  author = {Fiorella De Fiore and Alexis Maurin and Andrej Mijakovic and Damiano Sandri},
  date   = {2025},
  title  = {Monetary policy in the news: The FOMC’s media coverage and inflation expectations},
  url    = {https://cepr.org/voxeu/columns/monetary-policy-news-fomcs-media-coverage-and-inflation-expectations},
  urldate = {2025-06-02},

}

@inproceedings{Wei2022,
author = {Wei, Jason and Wang, Xuezhi and Schuurmans, Dale and Bosma, Maarten and Ichter, Brian and Xia, Fei and Chi, Ed H. and Le, Quoc V. and Zhou, Denny},
title = {Chain-of-thought prompting elicits reasoning in large language models},
year = {2022},
isbn = {9781713871088},
publisher = {Curran Associates Inc.},
booktitle = {Proceedings of the 36th International Conference on Neural Information Processing Systems},
articleno = {1800},
numpages = {14},
doi={10.5555/3600270.3602070},
series = {NIPS '22}
}

@techreport{Gehring2023,
  author = {Kai Gehring and Matteo Grigoletto},
  date   = {2023},
  title  = {Analyzing Climate Change Policy Narratives with the Character-Role Narrative Framework},
  type = {CESifo Working Paper},
  number = {10429},
  doi={10.2139/ssrn.4456361}
}

@Article{Han2024,
  author       = {Han, Di and Guo, Wei and Chen, Han and Wang, Bocheng and Guo, Zikun},
  date         = {2024},
  journaltitle = {International Review of Economics \& Finance},
  title        = {LEST: Large language models and spatio-temporal data analysis for enhanced Sino-US exchange rate forecasting},
  pages        = {103508},
  volume       = {96},
  publisher    = {Elsevier},
}

@inproceedings{POLAR2,
author = {Mathew, Binny and Sikdar, Sandipan and Lemmerich, Florian and Strohmaier, Markus},
title = {The POLAR Framework: Polar Opposites Enable Interpretability of Pre-Trained Word Embeddings},
year = {2020},
isbn = {9781450370233},
publisher = {Association for Computing Machinery},
address = {New York, NY, USA},
doi = {10.1145/3366423.3380227},
booktitle = {Proceedings of The Web Conference 2020},
pages = {1548–1558},
numpages = {11},
location = {Taipei, Taiwan},
series = {WWW '20}
}

@inproceedings{lex2sent,
    title = "{L}ex2{S}ent: A bagging approach to unsupervised sentiment analysis",
    author = "Lange, Kai-Robin  and
      Rieger, Jonas  and
      Jentsch, Carsten",
    booktitle = "Proceedings of the 20th Conference on Natural Language Processing (KONVENS 2024)",
    year = "2024",
    publisher = "Association for Computational Linguistics",
    url = "https://aclanthology.org/2024.konvens-main.28/",
    pages = "281--291",
}

@Article{Korinek2023,
  author       = {Korinek, Anton},
  date         = {2023},
  journaltitle = {Journal of Economic Literature},
  title        = {Generative AI for Economic Research: Use Cases and Implications for Economists},
  doi          = {10.1257/jel.20231736},
  issn         = {0022-0515},
  number       = {4},
  pages        = {1281--1317},
  volume       = {61},
  publisher    = {American Economic Association},
}

@Article{Jeong2025,
  author       = {Jeong, Minhyuk and Ahn, Kwangwon},
  date         = {2025},
  journaltitle = {Energy Economics},
  title        = {Energy organization sentiment and oil return forecast},
  pages        = {108105},
  volume       = {141},
  publisher    = {Elsevier},
}

@Article{Huang2023a,
  author       = {Huang, Lei and Yu, Weijiang and Ma, Weitao and Zhong, Weihong and Feng, Zhangyin and Wang, Haotian and Chen, Qianglong and Peng, Weihua and Feng, Xiaocheng and Qin, Bing and Liu, Ting},
  date         = {2023-11-09},
  journaltitle = {ACM Transactions on Information Systems},
  title        = {A Survey on Hallucination in Large Language Models: Principles, Taxonomy, Challenges, and Open Questions},
  doi          = {10.1145/3703155},
  eprint       = {2311.05232},
  eprintclass  = {cs.CL},
  eprinttype   = {arXiv},
  issn         = {1558-2868},
  number       = {2},
  pages        = {1--55},
  volume       = {43},
  copyright    = {arXiv.org perpetual, non-exclusive license},
  month        = jan,
  publisher    = {Association for Computing Machinery (ACM)},
  year         = {2023},
}

@inproceedings{zhang2021,
    title = "Trading Off Diversity and Quality in Natural Language Generation",
    author = "Zhang, Hugh  and
      Duckworth, Daniel  and
      Ippolito, Daphne  and
      Neelakantan, Arvind",
    editor = "Belz, Anya  and
      Agarwal, Shubham  and
      Graham, Yvette  and
      Reiter, Ehud  and
      Shimorina, Anastasia",
    booktitle = "Proceedings of the Workshop on Human Evaluation of NLP Systems (HumEval)",
    year = "2021",
    address = "Online",
    publisher = "Association for Computational Linguistics",
    url = "https://aclanthology.org/2021.humeval-1.3/",
    pages = "25--33",
}

@techreport{yang2025promptsdontsayunderstanding,
      title={What Prompts Don't Say: Understanding and Managing Underspecification in LLM Prompts}, 
      author={Chenyang Yang and Yike Shi and Qianou Ma and Michael Xieyang Liu and Christian Kästner and Tongshuang Wu},
      year={2025},
      eprint={2505.13360},
      archivePrefix={arXiv},
      primaryClass={cs.CL},
      url={https://arxiv.org/abs/2505.13360}, 
}

%\newpage
\section*{Appendix}
\renewcommand{\thesubsection}{\Alph{subsection}}

\subsection{Codebook}
\label{CodeBook}

\makeatletter
\newenvironment{tightverblines_codebuch}[1][0pt]{%
  \begingroup
  \renewcommand\section{\@startsection{section}{1}{\z@}%
    {3.5ex \@plus 1ex \@minus .2ex}%
    {1ex \@plus .3ex}%
    {\normalfont\small\bfseries}}
  \renewcommand\subsection{\@startsection{subsection}{2}{\z@}%
    {3.5ex \@plus 1ex \@minus .2ex}%
    {1ex \@plus .3ex}%
    {\normalfont\footnotesize\bfseries}}
  \renewcommand\subsubsection{\@startsection{subsubsection}{3}{\z@}%
    {3.5ex \@plus 1ex \@minus .2ex}%
    {1ex \@plus .3ex}%
    {\normalfont\footnotesize\bfseries\itshape}}

  \ttfamily
  \obeylines
  \footnotesize
  \setlength{\parskip}{#1}
  \setlength{\parindent}{0pt}

  % Kleine Item-Abstände
  \setlist[itemize]{topsep=2pt, partopsep=0pt, itemsep=1pt, parsep=0pt}
  \setlist[enumerate]{topsep=2pt, partopsep=0pt, itemsep=1pt, parsep=0pt}
}{%
  \endgroup
}
\makeatother

\begin{codebookbox}
\begin{tightverblines_codebuch}[0ex]
\section*{1. Basic Idea}

The goal of this coding task is to extract all narratives from the provided texts as accurately and comprehensively as possible. The texts are excerpts from newspaper articles published in the NYT or the Wall Street Journal.\footnote{Since the texts are excerpts, some passages may seem out of context. In most cases, however, the meaning should still be clear.}

\hspace{0.6em}
\textit{We define a narrative as a causal connection between two consecutive events.}

\hspace{0.6em}
\textbf{Simple examples of a narrative:}
\begin{itemize}
    \item Russia's war on Ukraine -- causes -- People leaving Ukraine
    \item rising prices -- causes -- lifting of global bond yields
    \item rapid money-supply growth -- is caused by -- influx of hard currency
\end{itemize}

\hspace{0.6em}
These narratives are sometimes explicitly mentioned in the text and easy to identify. Often, however, they require some interpretation and abstraction. That is why it is crucial for our research that multiple coders read and annotate the same texts. This ensures that results are not biased by individual interpretations and that the extracted narratives are recognizable by other coders as well.

\section*{2. Target Format}

We aim to train a language model to reduce all narratives in a text to \textbf{one of two possible formats}:

\begin{itemize}
    \item Format A: \texttt{event 1 -- causes -- event 2}
    \item Format B: \texttt{event 1 -- is caused by -- event 2}
\end{itemize}

\hspace{0.6em}
Each narrative must follow one of these formats. The distinction between A and B preserves the order of events as they appear in the original text. The event mentioned first must be \texttt{event 1} in your coded narrative.

\hspace{0.6em}
\textbf{Examples:}
\begin{itemize}
    \item \textit{Source text:} The Titanic struck an iceberg on 14 April 1912. As a result, the Titanic sank.

    \texttt{the Titanic struck an iceberg -- causes -- the Titanic sank}

    \item \textit{Source text:} People have a hard time finding jobs because the Fed keeps interest rates high.

    \texttt{people have a hard time finding jobs -- is caused by -- the Fed keeps interest rates high}
\end{itemize}

\hspace{0.6em}
These examples show: The narrative does not have to be a coherent sentence. The task is to reduce running text into a "bullet point" format that strips away all information not part of the causal chain.

\section*{3. Events}

\subsection*{3.1 Types of Events}

Events can be categorized into \textbf{events and activities} as well as \textbf{states and circumstances}.
\begin{itemize}
    \item ``the collapse of the Berlin Wall in 1989'' (event)
    \item ``Mexico grew rapidly'' (activity)
    \item ``Trump signed the bill into law'' (activity)
    \item ``inflation'' (state)
    \item ``lack of competitive innovation'' (circumstance)
\end{itemize}

\hspace{0.6em}
States and circumstances can also include \textbf{properties} of people or other entities:
\begin{itemize}
    \item ``Stubborn Indian inflation''
    \item ``The prime minister's arrogance''
\end{itemize}

\hspace{0.6em}
Events and activities can also include \textbf{future events or plans}.
\begin{itemize}
    \item ``The ECB will probably raise rates by next year'' (plan)
    \item ``There might be more earthquakes in Germany in the future'' (future event)
\end{itemize}

\hspace{0.6em}
\textbf{Policy measures} also count as events:
\begin{itemize}
    \item ``The Schuldenbremse''
    \item ``The Affordable Care Act''
\end{itemize}

\subsection*{3.2 Unchanged Coding of Events}

Events should generally be coded in the \textbf{exact form used in the source text}. The closer the extracted narrative is to the source text, the faster the model will learn.

\hspace{0.6em}
Specifically:
\begin{itemize}
    \item Do not use synonyms for terms used in the text
    \item Keep the same verb tense (present stays present, past stays past, etc.)
    \item Do not abstract semantically or economically (e.g., do not interpret ``rising prices'' as ``inflation'')
    \item \textbf{Exception:} Explanatory clauses (e.g., appositions) can usually be left out (see 3.5)
\end{itemize}

\subsection*{3.3 Coreference Resolution}

An exception to 3.2 applies when entities (e.g., people, countries) are only indirectly mentioned.
\hspace{0.6em}
Example:
\textit{``She talked a lot about North Korea. 'The country is really poor, therefore many citizens suffer from hunger', she said.''}

--> Instead of coding ``The country'', include the resolved entity:

\begin{itemize}
    \item \texttt{\{The country|North Korea\} is really poor -- causes -- citizens suffer from hunger}
\end{itemize}

\hspace{0.6em}
Similarly:

\textit{``President Biden discussed the matter today. He downplayed the probability of a government shutdown which has calmed the markets.''}

\begin{itemize}
    \item \texttt{\{He|President Biden\} downplayed the probability of a government shutdown -- causes -- the markets were calmed}
\end{itemize}

\subsection*{3.4 Opinions}

Sometimes a causal link is part of a third-party opinion or assessment. In most cases, the narrative refers to the \textbf{content} of the opinion, not the act of expressing it.

\hspace{0.6em}
\textbf{Example:}

\textit{``Analysts believe that bond yields might rise even further which would put even more strain on pension funds.''}

\begin{itemize}
    \item \texttt{bond yields might rise even further -- causes -- more strain on pension funds}
\end{itemize}

\hspace{0.6em}
But in some cases, the act of expression itself is causal:

\hspace{0.6em}
\textit{``ECB president Lagarde made it clear that no further rate raises were on the horizon. Markets responded calmly.''}

\begin{itemize}
    \item \texttt{ECB president Lagarde made it clear that no further rate raises were on the horizon -- causes -- Markets responded calmly}
\end{itemize}

\subsection*{3.5 Parentheses and Side Clauses}

Explanatory side information set off by commas, dashes, or parentheses should be left out unless essential for understanding the narrative.

\hspace{0.6em}
\textbf{Example:}

\textit{``The highest inflation in decades and the war in Ukraine -- the first major military conflict in Europe in over 30 years -- raised concerns...''}

\begin{itemize}
    \item highest inflation in decades -- causes -- concerns about a possible recession
    \item war in Ukraine -- causes -- concerns about a possible recession
\end{itemize}

\subsection*{3.6 Chained Events}

Some statements include multiple events connected by ``and'' or ``as well as''. These should be split into separate narratives.

\hspace{0.6em}
\textbf{Example:}

\textit{``She wants you to forget that federal spending contributed to soaring prices, as well as to the labor shortages across the economy.''}

\begin{itemize}
    \item federal spending -- causes -- soaring prices
    \item federal spending -- causes -- labor shortages across the economy
\end{itemize}

\section*{4. Only Positive Causal Links}

Please code only \textbf{positive} causal relationships. Ignore negative causality.

\hspace{0.6em}
\textbf{Example:}

\texttt{Putin invading Ukraine -- does not cause -- President Zelensky fleeing Ukraine} \textit{(do not code)}

\section*{5. Multiple Narratives per Text}

A document may contain multiple narratives. Please write \textbf{each narrative on a new line}. Do not use enumerations.

\hspace{0.6em}
\textbf{Correct:}
- the prices for cucumbers rose by 100\% this month -- causes -- people stop buying cucumbers
- huge money supply -- causes -- prices for real estate go up

\hspace{0.6em}
\textbf{Wrong:}
1. the prices for cucumbers rose ... | 2. huge money supply ...

\section*{6. Thematic Scope}

\subsection*{6.1 Explicit Economic Reference}

Please only code narratives that have an explicit economic relevance. You can ignore all other narratives. If you are unsure, write down the narrative and add a note (uncertain).

\hspace{0.6em}
\textbf{Examples:}

\begin{itemize}
    \item \texttt{Putin invading Ukraine -- causes -- rising energy prices} \textit{(code)}
    \item \texttt{Putin invading Ukraine -- causes -- enormous military support by the United States} \textit{(do not code)}
\end{itemize}

\subsection*{6.2 Focus on Inflation Narratives}

We are primarily interested in \textbf{inflation narratives}. Therefore, please mark all narratives \textbf{without} explicit reference to inflation or price changes with an ``(x)''. As a reminder: narratives that have no economic relevance at all should not be coded in the first place.

\hspace{0.6em}
\textbf{Examples:}
\begin{itemize}
    \item \texttt{Putin invading Ukraine -- causes -- disruption of agricultural supply chains (x)} \textit{(code)}
    \item \texttt{Putin invading Ukraine -- causes -- gas price increases in Germany} \textit{(code)}
    \item \texttt{Putin invading Ukraine -- causes -- enormous military support by the United States} \textit{(do not code)}
\end{itemize}

\hspace{0.6em}
\textbf{The example from 3.6 should therefore be:}
\begin{itemize}
    \item \texttt{federal spending -- causes -- soaring prices}
    \item \texttt{federal spending -- causes -- labor shortages across the economy (x)}
\end{itemize}

\section*{7. Preserve Content}

This point is similar to point 3.2. Sometimes, there is a strong temptation to omit parts of a sentence in order to make the narrative more pointed. In some cases, that's acceptable; however, the rule should be to include all parts of the sentence that are necessary to preserve the core message of the narrative. Please preserve all essential content in the narrative.

\hspace{0.6em}
\textbf{Example 1:}
\begin{itemize}
    \item \texttt{wage inflation going up -- causes -- the risk of yields on longer-dated bonds rising at a faster pace compared to those on shorter-dated notes} (correct)
    \item \texttt{wage inflation going up -- causes -- longer-dated bonds rising faster than shorter-dated notes} (incorrect)
\end{itemize}

\hspace{0.6em}
\textbf{Example 2:}
\begin{itemize}
    \item \texttt{The highest inflation in decades -- causes -- concerns about a possible recession} (correct)
    \item \texttt{inflation -- causes -- recession} (incorrect)
\end{itemize}

\section*{8. Edge Cases}

If you are unsure whether something really constitutes a narrative or not, code it anyway to be on the safe side. You can later compare all such cases with the other coders or with us.

\hspace{0.6em}
\textbf{Example 1:}
"Semi"-causal cues (such as "as", "suggesting", ...) can be interpreted as either temporal or causal.

\hspace{0.6em}
\textit{John Sicher, editor and publisher of Beverage Digest, said that soft-drink sales have shown some improvement this year, estimating that they rose about 3.5\% in January as price increases stabilized.}

\hspace{0.6em}
In such cases code the narrative just to be safe, and compare your results with other coders afterward.

\hspace{0.6em}
\textbf{Example 2:}
If you interpret the "as" in the example above as a causal cue, a second ambiguity arises: you could shorten the passage in two ways:

\hspace{0.6em}
\textit{John Sicher, editor and publisher of Beverage Digest, said that soft-drink sales have shown some improvement this year, estimating that they rose about 3.5\% in January as price increases stabilized.}

\begin{itemize}
    \item price increases stabilized – causes – soft-drink sales showing some improvement (x)
    \item price increases stabilized – causes – soft-drink sales rose about 3.5\% in January (x)
\end{itemize}

\hspace{0.6em}
Again, if you’re unsure which version to choose, write down both to be on the safe side.

\hspace{0.6em}
If unsure, include both options.

\end{tightverblines_codebuch}
\end{codebookbox}

\newpage
\subsection{Annotated Examples}
\label{AnnotatedExample}

\begin{codebookbox}
\begin{tightverblines_codebuch}[0ex]

\textit{2004-05-27 | WSJ | China Targets Auto Loans, Speculators}
\vspace{10pt}
At the same time, overinvestment in the industry has led to rising car inventories. GM says its own inventory levels are healthy and for some models it has waiting lists. Separately, China's foreign-exchange regulator announced plans to slow an influx of hard currency that is fueling rapid money-supply growth and pushing up inflation. Speculators are betting that Beijing will revalue its currency, the yuan, and are parking money in Chinese banks to take advantage of interest rates higher than those in the U.S.The new rules also aim to reduce a run-up in foreign debt, as some borrowers switch to foreign-currency loans that are cheaper than yuan-denominated loans. As part of the tighter measures to come into force on July 1, foreign-invested companies must state a specific use for any funds greater than \$200,000 entering China, said a statement posted on the State Administration of Foreign Exchange's Web site.

\subsection*{Narratives}

\begin{itemize}
    \item overinvestment in the industry - causes - rising car inventories (x)
    \item an influx of hard currency - causes - rapid money-supply growth
    \item an influx of hard currency - causes - pushing up inflation
    \item run-up in foreign debt - is caused by - some borrowers switch to foreign-currency loans that are cheaper than yuan-denominated loans
    \item The new rules - causes - reducing a run-up in foreign debt. (x)
\end{itemize}

\subsection*{Model Output}

\begin{itemize}
    \item overinvestment in the car industry - causes - rising car inventories
    \item an influx of hard currency - causes - rapid money-supply growth
    \item rapid money-supply growth - causes - pushing up inflation
    \item Speculators are betting that Beijing will revalue its currency, the yuan - causes - speculators parking money in Chinese banks to take advantage of interest rates higher than those in the U.S.
    \item foreign-currency loans are cheaper than yuan-denominated loans - causes - some borrowers switch to foreign-currency loans
\end{itemize}

\end{tightverblines_codebuch}
\end{codebookbox}

\begin{codebookbox}
\begin{tightverblines_codebuch}[0ex]

\textit{2023-03-03 | WSJ | Business News: Bud Brewer's Volume Fell as Prices Jumped}
\vspace{10pt}
And people are eating and drinking more at home, a setting that tends to favor beer over other drinks, he said\". Beer remains resilient,\" Mr. Doukeris said. \"I'll never say that beer is immune to the inflation and everything that is happening out there, but [it] remains a very resilient category\". The brewer has for years seen its mainstream brands Bud and Bud Light decline, partly amid a broader shift by American drinkers from beer toward wine and spirits. In response, it has worked to increase its exposure to premium offerings such as Michelob Ultra and Stella Artois.

\subsection*{Narratives}

\begin{itemize}
    \item people are eating and drinking more at home, a setting that tends to favor beer over other drinks - causes - Beer remains resilient (x)
    \item Bud and Bud Light decline - is caused by - a broader shift by American drinkers from beer toward wine and spirits (x)
    \item Bud and Bud Light decline - causes - \{it|the brewer\} has worked to increase its exposure to premium offerings such as Michelob Ultra and Stella Artois (x)
\end{itemize}

\subsection*{Model Output}

\begin{itemize}
    \item people are eating and drinking more at home - causes - beer being favored over other drinks
    \item a broader shift by American drinkers from beer toward wine and spirits - causes - Bud and Bud Light decline
    \item Bud and Bud Light decline - causes - the brewer worked to increase its exposure to premium offerings such as Michelob Ultra and Stella Artois
\end{itemize}

\end{tightverblines_codebuch}
\end{codebookbox}

\begin{codebookbox}
\begin{tightverblines_codebuch}[0ex]

\textit{2023-03-03 | NYT | Market Gains After 3 Days of Losses}
\vspace{10pt}
Shares of Salesforce soared 11.5 percent for one of the market's biggest gains after it topped forecasts for profit and revenue last quarter. It also gave a stronger-than-expected forecast for upcoming results. Expectations have been coming down recently for profits at big U.S. companies given still-high inflation and interest rates. But several joined Salesforce in rising Thursday after posting encouraging results. Macy's rose 11.1 percent after reporting stronger profit and revenue for the holidays than analysts expected.

\subsection*{Narratives}

\begin{itemize}
    \item Shares of Salesforce soared 11.5 percent for one of the market's biggest gains - is caused by - \{it|Salesforce\} topped forecasts for profit and revenue last quarter. (x)
    \item Shares of Salesforce soared 11.5 percent for one of the market's biggest gains - is caused by - \{it|Salesforce\} gave a stronger-than-expected forecast for upcoming results (x)
    \item \{it|Salesforce\} gave a stronger-than-expected forecast for upcoming results (x) - is caused by - \{it|Salesforce\} topped forecasts for profit and revenue last quarter. (x)
    \item Expectations have been coming down recently for profits at big U.S. companies - is caused by - still-high inflation
    \item Expectations have been coming down recently for profits at big U.S. companies - is caused by - still-high interest rates
    \item \{Several|U.S. companies\} joined Salesforce in rising Thursday - is caused by - \{U.S. companies\} posting encouraging results (x)
    \item Macy's rose 11.1 percent - is caused by - reporting stronger profit and revenue for the holidays than analysts expected (x)
\end{itemize}

\subsection*{Model Output}

\begin{itemize}
    \item Shares of Salesforce soared 11.5 percent for one of the market's biggest gains - is caused by - Salesforce topped forecasts for profit and revenue last quarter
    \item Salesforce gave a stronger-than-expected forecast for upcoming results - causes - Shares of Salesforce soared 11.5 percent for one of the market's biggest gains
    \item Macy's rose 11.1 percent - is caused by - Macy's reported stronger profit and revenue for the holidays than analysts expected
\end{itemize}

\end{tightverblines_codebuch}
\end{codebookbox}

\subsection{Prompt}
\label{Prompt}

\newenvironment{tightverblines}[1][0pt]{%
  \ttfamily
  \obeylines
  \footnotesize
  \setlength{\parskip}{#1} % allows you to set spacing
  \setlength{\parindent}{0pt}
}{}

\begin{codebookbox}
\begin{tightverblines}[0ex]
\# Codebook:
\vspace{3ex}
\#\# Basic Idea:
I will provide you with an excerpt from newspaper articles that have appeared in the New York Times or The Wall Street Journal.
Your task is to extract all economic narratives that occur in the excerpt. This task is a well-defined multi-step extraction process. The term "narrative" is frequently used with different meanings and in a variety of contexts. The multi-step extraction process is designed to make sure that your work respects a very specific definition of economic narratives. For each step that outlined below, it is imperative that you only make the changes that are indicated. I will provide you with this definition next.
\vspace{3ex}
\#\# Definition:
An economic narrative consists of exactly two events and a causal connection that is asserted between those events.
\vspace{3ex}
\#\# Event structures:
In the terminology of causal inference, the definition corresponds to finding the "direct causal effect" of Event A on Event B. The causal pathways you encounter in the excerpts may be more complex. Therefore, in addition to direct causal effects, find and deconstruct the following causal structures as follows:
1) Fork: If Event A is said to be a common cause of Events B and C, code both pathways as separate narratives.
2) Chain: If Event B is said to be a mediator between Event A and Event C, also code both pathways as separate narratives.
\vspace{3ex}
\#\# Rules:
You also have to follow a couple of hard-and-fast rules:
1) Retain the order of the two events from the source text at all times. The event that appears first in the text must be coded as "Event A", the event that appears second must be coded as "Event B".
2) When you state the narrative in its "Target Form", always represent the causal connection by using one of the following phrasings: i) "causes": Use this causal connector when Event A is the cause and Event B is an effect of Event A. ii) "is caused by": Use this causal connector when Event B is the cause and Event A is an effect of Event B.
\vspace{3ex}
\#\# Target Forms:
\{
"Event A": "[...]",
"causal connector": "causes",
"Event B": "[...]"
\}
\{
"Event A": "[...]",
"causal connector": "caused by",
"Event B": "[...]"
\}
\vspace{3ex}
\#\# Extraction Process:
1) "Focused Excerpt": The excerpts vary in length and in their narrative density. Parts of each excerpt may obviously not contain any narrative. To focus on relevant parts of the excerpt, start by repeating it, but leaving out the sentences that you are sure no narrative appears in.
2) "Sequence of Interest": Go through the "Focused Excerpt" and state the first narrative sequence you find, that is a sequence that contains two events and what you consider to be a causal connection between them.
3) "Causal Restatement": Based on the current "Sequence of Interest", restate the narrative in the appropriate target form. Make sure to repeat the events verbatim without rephrasing.
4) "Coreference Resolution": If necessary, rephrase one or both events to clearly identify the entities that occur in the narrative. For the most part, this will involve replacing personal pronouns with the entity itself.
5) "Event Rephrasing": If necessary, rephrase one or both events again so that the narrative uses correct language.

After every narrative, continue by circling back to 1) and restate the "Focused Excerpt". Do so even if you did not previously detect more narratives in it. Use it to refocus your attention and double-check if more economic narratives occur in the excerpt.
\vspace{3ex}
\# Example:
\vspace{3ex}
\#\# Excerpt:
\textit{"[...]"}
\{
"Focused Excerpt": \textit{"[...]"},
"Sequence of interest": \textit{"[...]"},
"Causal Restatement":
\{
"Event A": \textit{"[...]"},
"causal connector": "causes",
"Event B": \textit{"[...]"}
\},
"Coreference Resolution":
\{
"Event A": \textit{"[...]"},
"causal connector": "causes",
"Event B": \textit{"[...]"}
\},
"Event Rephrasing":
\{
"Event A": \textit{"[...]"},
"causal connector": "causes",
"Event B": \textit{"[...]"}
\}
\},
\{
"Focused Excerpt": \textit{[...]},
\textit{[...]}
\}
\vspace{3ex}
\# Your Work
\#\# Excerpt:
\textit{"[...]"}
\end{tightverblines}
\end{codebookbox}

\end{document}